\documentclass[numreferences]{kluwer}

\usepackage{epsfig}

\newcommand{\ec}[0]{\epsilon_C}
\newcommand{\el}[0]{\epsilon_L}
\newcommand{\er}[0]{\epsilon_R}
\newcommand{\dc}[0]{\( D_C \ \)}
\newcommand{\dr}[0]{\( D_R \ \)}
\newcommand{\dl}[0]{\( D_L \ \)}

\begin{document}
\typeout{^^J*** Kluwer Academic Publishers - prepress department^^J***
    Documentation for book editors using the Kluwer class files}
\begin{article}
\begin{opening}
\title{Shot noise for entangled and spin-polarized electrons}
\author{J. C. \surname{Egues}\thanks{Permanent address:
Department of Physics and Informatics, University of S\~ao Paulo
at S\~ao Carlos, 13560-970 S\~ao Carlos/SP, Brazil.}, P. Recher,
D. S. Saraga, V. N. Golovach, G. Burkard, E. V. Sukhorukov, and D.
Loss} \runningauthor{Egues, Recher, Saraga, Golovach, Burkard,
Sukhorukov, and Loss } \runningtitle{Shot noise for entangled and
spin-polarized electrons} \institute{Department of Physics and
Astronomy, University of Basel, Klingelbergstrasse 82, CH-4056
Basel, Switzerland}
\begin{abstract}
We review our recent contributions on shot noise for entangled
electrons and spin-polarized currents in novel mesoscopic
geometries. We first discuss some of our recent proposals for
electron entanglers involving a superconductor coupled to a double
dot in the Coulomb blockade regime, a superconductor
tunnel-coupled to Luttinger-liquid leads, and a triple-dot setup
coupled to Fermi leads. We briefly survey some of the available
possibilities for spin-polarized sources. We use the scattering
approach to calculate current and shot noise for spin-polarized
currents and entangled/unentangled electron pairs in a novel
beam-splitter geometry with a \textit{local} Rashba spin-orbit
(s-o) interaction in the incoming leads. For single-moded incoming
leads, we find \textit{continuous} bunching and antibunching
behaviors for the \textit{entangled} pairs -- triplet and singlet
-- as a function of the Rashba rotation angle. In addition, we
find that unentangled triplets and the entangled one exhibit
distinct shot noise; this should allow their identification via
noise measurements. Shot noise for spin-polarized currents shows
sizable oscillations as a function of the Rashba phase. This
happens only for electrons injected perpendicular to the Rashba
rotation axis; spin-polarized carriers along the Rashba axis are
noiseless. The Rashba coupling constant $\alpha$ is directly
related to the Fano factor and could be extracted via noise
measurements. For incoming leads with s-o induced
interband-coupled channels, we find an additional spin rotation
for electrons with energies near the crossing of the bands where
interband coupling is relevant. This gives rise to an additional
modulation of the noise for both electron pairs and spin-polarized
currents. Finally, we briefly discuss shot noise for a double dot
near the Kondo regime.
\end{abstract}


\end{opening}

\section{Introduction}

Fluctuations of the current away from its average usually contain
supplementary information, not provided by average-current
measurements alone. This is particularly true in the non-linear
response regime where these quantities are not related via the
fluctuation-dissipation theorem. At zero temperature,
non-equilibrium current noise is due to the discreteness of the
electron charge and is termed shot noise. This dynamic noise was
first investigated by Schottky in connection with thermionic
emission \cite{schottky}. Quantum shot noise has reached its come
of age in the past decade or so and constitutes now an
indispensable tool to probe mesoscopic transport
\cite{blanter-buttiker}; in particular, the role of fundamental
correlations such as those imposed by quantum statistics.

More recently, shot noise has been investigated in connection with
transport of entangled \cite{de}--\cite{prl-cgd} and spin
polarized electrons \cite{gde}, \cite{prl-cgd}--\cite{sej} and has
proved to be a useful probe for both entanglement and
spin-polarized transport. Entanglement \cite{entanglement} is
perhaps one of the most intriguing features of quantum mechanics
since it involves the concept of non-locality. Two-particle
entanglement is the simplest conceivable form of entanglement.
Yet, these Einstein-Podolsky-Rosen (EPR) pairs play a fundamental
role in potentially revolutionary implementations of quantum
computation, communication, and information processing
\cite{als-review}. In this context, such a pair represents two
qubits in an entangled state. The generation and detection of EPR
pairs of photons has already been accomplished. On the other hand,
research involving two-particle entanglement of massive particles
(e.g. electrons) in a solid-state matrix is still in its infancy,
with a few proposals for its physical implementation; some of
these involve quantum-dot setups as sources of mobile
spin-entangled electrons \cite{gde}, \cite{P3RSL}, \cite{Sar02}.
Spin-polarized transport \cite{ohno-molen}, \cite{egues},
\cite{patrik}, on the other hand, is a crucial ingredient in
semiconductor spintronics where the spin (and/or possibly the
charge) of the carriers play the relevant role in a device
operation. To date, robust spin injection has been achieved in
Mn-based semiconductor layers (\textit{pin} diode structures)
\cite{ohno-molen}. High-efficiency spin injection in other
semiconductor systems such as hybrid ferromagnetic/semiconductor
junctions is still challenging.

It is clear that the ability to create, transport, coherently
manipulate, and detect entangled electrons and spin-polarized
currents in mesoscopic systems is highly desirable. Here we review
some of our recent works \cite{gde}, \cite{jsc-cgd},
\cite{prl-cgd}, \cite{P3RSL}, \cite{Sar02}, \cite{P8RL},
\cite{GolovachLoss} addressing some of these issues and others in
connection with noise. Shot noise provides an additional probe in
these novel transport settings. We first address the production of
mobile entangled electron pairs (Sec. \ref{entanglers}). We
discuss three proposals involving a superconductor coupled to two
dots \cite{P3RSL}, a superconductor coupled to Luttinger-liquid
leads \cite{P8RL}, and a triple-dot arrangement \cite{Sar02}. Our
detailed analysis of these ``entanglers'' does not reveal any
intrinsic limitation to their experimental feasibility. We also
mention some of the available sources of spin-polarized electrons
(Sec. \ref{spinfilters}). Ballistic spin filtering with
spin-selective semimagnetic tunnel barriers \cite{egues} and
quantum dots as spin filters \cite{patrik} are also briefly
discussed.

\begin{figure}[th]
\begin{center}
\epsfig{file=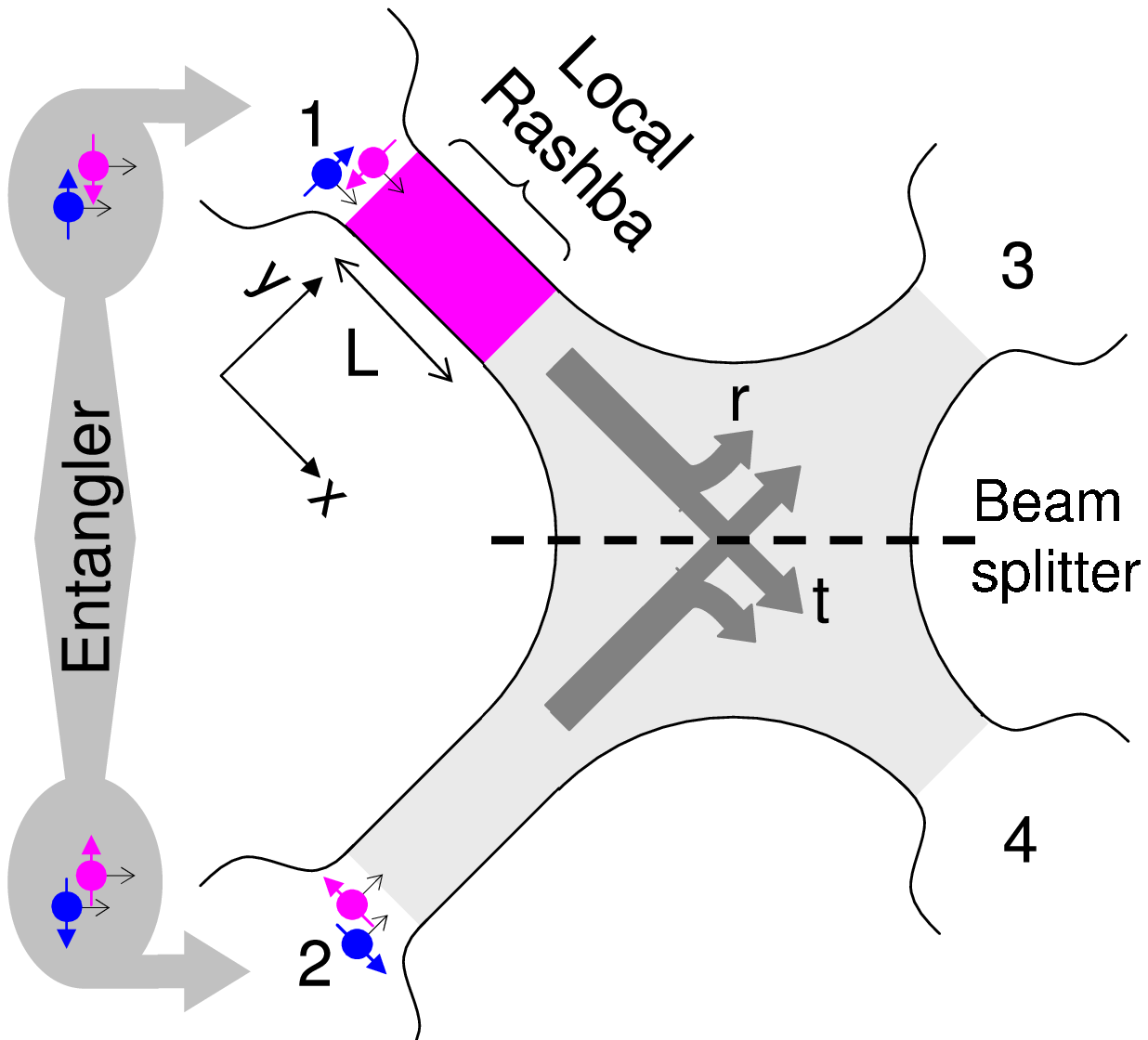, width=0.80\textwidth}
\end{center}
\caption{Novel electron beam-splitter geometry with a local Rashba
s-o interaction in lead 1. An entangler or a spin-polarized
electron source (not shown) inject either entangled pairs or
spin-polarized carriers into leads 1 and 2. The portion of the
entangled pairs (or the spin-polarized carriers) traversing lead 1
undergoes a Rashba-induced spin rotation. This \emph{continuously}
changes the symmetry of the \emph{spin} part of the pair wave
function. Adapted from Ref. \cite{prl-cgd} .} \label{bs-fig1}
\end{figure}

We investigate transport of entangled and spin polarized electrons
in a beam-splitter (four-port) configuration \cite{bs},
\cite{henny} with a local Rashba spin-orbit interaction in the
incoming leads \cite{feve}, Fig. (\ref{bs-fig1}). A local Rashba
term provides a convenient way to coherently spin-rotate electrons
as they traverse quasi one-dimensional channels, as was first
pointed out by Datta and Das \cite{datta-das}.  Within the
scattering formalism \cite{blanter-buttiker}, we calculate shot
noise for both entangled and spin-polarized electrons.

For entangled electrons, shot noise is particularly relevant as a
probe for fundamental two-particle interference. More
specifically, shot noise (charge noise) directly probes the
orbital symmetry of the EPR pair wave function. However, the
symmetry of the orbital degree of freedom (``the charge'') is
intrinsically tied to that of the spin part of the pair wave
function via the Pauli principle. That is, the total electron-pair
wave function is antisymmetric thus imposing a fundamental
connection between the spin and orbital parts of the pair wave
function. Hence charge noise measurements probe in fact the spin
symmetry of the pair. Moreover, if one can alter the spin state of
the pair (say, via some proper coherent spin rotation) this will
definitely influence shot-noise measurements. This is precisely
what we find here for singlet and triplet pairs.

The coherent local Rashba spin rotation in one of the incoming
leads of our setup, continuously alters the (spin) symmetry of the
pair wave function thus giving rise to sizable shot noise
oscillations as a function of the Rashba phase. Noise measurements
in our novel beam-splitter should allow one to distinguish
entangled triplets from singlets and entangled triplets from the
unentangled ones, through their Rashba phase. Entangled pairs
display continuous bunching/antibunching behavior. In addition,
triplets (entangled or not) defined along different quantization
axes (\textit{x}, \textit{y}, or \textit{z}) exhibit distinctive
noise, thus allowing the detection of their spin polarization via
charge noise measurements.

Shot noise for spin-polarized currents also probes effects imposed
by the Pauli principle through the Fermi functions in the leads.
These currents also exhibit Rashba-induced oscillations for spin
polarizations perpendicular to the Rashba rotation axis. We find
zero shot noise for spin-polarized carriers with polarizations
along the Rashba axis and for unpolarized injection. Moreover, the
Rashba-induced modulations of the Fano factor for both entangled
and spin-polarized electrons offer a direct way to extract the s-o
coupling constant via noise measurements.

We also consider incoming leads with two transverse channels. In
the presence of a weak s-o induced interchannel coupling, we find
an additional spin rotation due to the coherent transfer of
carriers between the coupled channels in lead 1. This extra
rotation gives rise to further modulation of the shot noise
characteristics for both entangled and spin-polarized currents;
this happens only for carriers with energies near the band
crossings in lead 1. Finally, we briefly discuss shot noise for
transport through a double dot near the Kondo regime
\cite{GolovachLoss}.

\section{Sources of mobile spin-entangled electrons}
\label{entanglers}

A challenge in mesoscopic physics is the experimental realization
of an electron ``entangler'' -- a device creating mobile entangled
electrons which are spatially separated. Indeed, these are
essential for quantum communication schemes and experimental tests
of quantum non-locality with massive particles. First, one should
note that entanglement is rather the rule than the exception in
nature, as it arises naturally from Fermi statistics. For
instance, the ground state of a helium atom is the spin singlet
$|\!\!\uparrow \downarrow \rangle - |\!\!\downarrow \uparrow
\rangle$. Similarly, one finds a singlet in the ground state of a
quantum dot with two electrons. These ``artificial atoms''
\cite{Kou97} are very attractive for manipulations at the single
electron level, as they possess tunable parameters and allow
coupling to mesoscopic leads -- contrary to real atoms. However,
such ``local'' entangled singlets are not readily useful for
quantum computation and communication, as these require control
over each individual electron as well as non-local correlations.
An improvement in this direction is given by two coupled quantum
dots with a single electron in each dot \cite{ld}, where the
spin-entangled electrons are already spatially separated by strong
on-site Coulomb repulsion (like in a hydrogen molecule). In this
setup, one could create mobile entangled electrons by
simultaneously lowering the tunnel barriers coupling each dot to
separate leads. Another natural source of spin entanglement can be
found in superconductors, as these contain Cooper pairs with
singlet spin wave functions. It was first shown in Ref.
\cite{P1CBL} how a non-local entangled state is created in two
uncoupled quantum dots when coupled to the same superconductor. In
a non-equilibrium situation, the Cooper pairs can be extracted to
normal leads by Andreev tunnelling, thus creating a flow of
entangled pairs
\cite{P3RSL},\cite{P8RL},\cite{P13BVBF}--\cite{Bou02}.

A crucial requirement for an entangler is to create {\em spatially
separated} entangled electrons; hence one must avoid whole
entangled pairs entering the same lead. As will be shown below,
energy conservation is an efficient mechanism for the suppression
of undesired channels. For this, interactions can play a decisive
role. For instance, one can use Coulomb repulsion in quantum dots
\cite{P3RSL},\cite{Sar02} or in Luttinger liquids
\cite{P8RL},\cite{P13BVBF}. Finally, we mention recent entangler
proposals using leads with narrow bandwidth \cite{P20Oliver}
and/or generic quantum interference effects \cite{Bos02}. In the
following, we discuss our proposals towards the realization of an
entangler that produces mobile non-local singlets
\cite{entang-triplet}. We set $\hbar=1$ in this section.

\subsection{Superconductor-based Electron entanglers}
\label{ssentanglers}

Here we envision a  {\em non-equilibrium} situation in which the
electrons of a Cooper pair tunnel coherently by means of an
Andreev tunnelling event from a SC to two separate normal leads,
one electron per lead. Due to an applied bias voltage, the
electron pairs can move into the leads thus giving rise to mobile
spin entanglement. Note that an (unentangled) single-particle
current is strongly suppressed by energy conservation as long as
both the temperature and the bias are  much smaller than the
superconducting gap. In the following we review two proposals
where we exploit the repulsive Coulomb charging energy between the
two spin-entangled electrons in order to separate them so that the
residual current in the leads is carried by non-local singlets. We
show that such entanglers meet all requirements for subsequent
detection of spin-entangled electrons via noise measurements
(charge measurement, see Secs. \ref{earlier-results} and
\ref{noise-rashba}) or via spin-projection measurements (Bell-type
measurement, see Sec. \ref{bell}).

\subsubsection{Andreev entangler with quantum dots}
\label{ssecAndreev}

\begin{figure}[h]
\centerline{\psfig{file=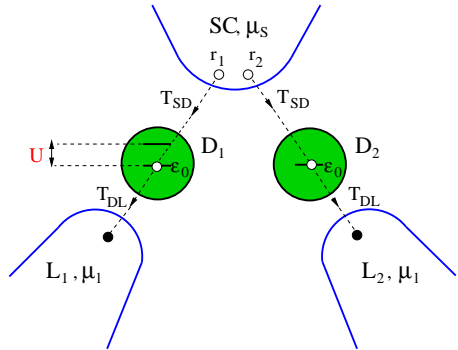,width=6cm}} \vspace{5mm}
\caption{The entangler setup. Two spin-entangled electrons forming
a Cooper pair tunnel with amplitude $T_{SD}$ from points ${\bf
r}_{1}$ and ${\bf r}_{2}$ of the superconductor, SC, to two dots,
$D_{1}$ and $D_{2}$, by means of Andreev tunnelling. The dots are
tunnel-coupled to normal Fermi liquid leads ${\rm L_{1}}$ and
${\rm L_{2}}$, with tunnelling amplitude $T_{DL}$. The
superconductor and leads are kept at chemical potentials $\mu_{S}$
and $\mu_{l}$, respectively. Adapted from \cite{P3RSL}.}
\label{spintabl}
\end{figure}

The proposed entangler setup (see Fig.~\ref{spintabl}) consists of
a SC with chemical potential $\mu_{S}$ which is weakly coupled to
two quantum dots (QDs) in the Coulomb blockade regime
\cite{Kou97}. These QDs are in turn  weakly coupled to outgoing
Fermi liquid leads, held at the same chemical potential $\mu_{l}$.
A bias voltage $\Delta\mu=\mu_{S}-\mu_{l}$ is applied between the
SC and the leads. The tunnelling amplitudes between the SC and the
dots, and dots and leads, are denoted by $T_{SD}$ and $T_{DL}$,
respectively (see Fig.~\ref{spintabl}).
The two intermediate QDs in the Coulomb blockade regime have
chemical potentials $\epsilon_{1}$ and $\epsilon_{2}$,
respectively. These can be tuned via external gate voltages, such
that the tunnelling of two electrons via different dots into
different leads is resonant for
$\epsilon_{1}+\epsilon_{2}=2\mu_{S}$ \cite{energyconservation}. As
it turns out \cite{P3RSL}, this two-particle resonance is
suppressed for the tunnelling of two electrons via the same dot
into the same lead by the on-site repulsion $U$ of the dots and/or
the superconducting gap $\Delta$. Next, we specify the parameter
regime of interest here in which the initial spin-entanglement of
a Cooper pair in the SC is successfully transported to the leads.

Besides the fact that single-electron tunnelling and tunnelling of
two electrons via the same dot should be excluded, we also have to
suppress transport of electrons which are already on the QD´s.
This could lead to effective spin-flips on the QD´s, which would
destroy the spin entanglement of the two electrons tunnelling into
the Fermi leads. A further source of unwanted spin-flips on the
QD´s is provided by its coupling to the Fermi liquid leads via
particle-hole excitations in the leads. The QDs can be treated
each as one localized spin-degenerate level as long as the mean
level spacing $\delta\epsilon$ of the dots exceeds both the bias
voltage $\Delta\mu$ and the temperature $k_{B}T$. In addition, we
require that each QD contains an even number of electrons with a
spin-singlet ground state. A more detailed analysis of such a
parameter regime is given in \cite{P3RSL} and is stated here
\begin{equation}
\label{regime} \Delta,U,
\delta\epsilon>\Delta\mu>\gamma_{l},k_{B}T,{\rm
and}\,\,\gamma_{l}>\gamma_{S}.
\end{equation}

In (\ref{regime}) the rates for tunnelling of an electron from the
SC to the QDs and from the QDs to the Fermi leads are given by
$\gamma_{S}=2\pi\nu_{S}|T_{SD}|^{2}$  and
$\gamma_{l}=2\pi\nu_{l}|T_{DL}|^{2}$, respectively,  with
$\nu_{S}$ and $\nu_{l}$ being the corresponding  electron density
of states per spin at the Fermi level. We consider asymmetric
barriers $\gamma_{l}>\gamma_{s}$ in order to exclude correlations
between subsequent Cooper pairs on the QDs. We work at the
particular interesting resonance $\epsilon_{1},\epsilon_{2}\simeq
\mu_{S}$, where the injection of the electrons into different
leads takes place at the same orbital energy. This is a crucial
requirement for the subsequent detection of entanglement via noise
\cite{gde,prl-cgd}. In this regime, we have calculated and
compared the stationary charge current of two spin-entangled
electrons for two competing transport channels in a T-matrix approach.\\

The ratio of the desired current for two electrons tunnelling into
{\em different} leads ($I_{1}$) to the unwanted current for two
electrons into the {\em same} lead ($I_{2}$) is~\cite{P3RSL}
\begin{equation}
\label{final} \frac{I_{1}}{I_{2}}= \frac{4{\cal E}^2}{\gamma^2}
\left[\frac{\sin(k_{F}\delta r)}{k_{F}\delta
r}\right]^2\,e^{-2\delta r/\pi\xi}, \quad\quad\quad
   \frac{1}{{\cal E}}=\frac{1}{\pi\Delta}+\frac{1}{U},
\end{equation}
where $\gamma = \gamma_1 + \gamma_2$. The current $I_1$ becomes
exponentially suppressed with increasing distance $\delta r=|{\bf
r}_1-{\bf r}_2|$ between the tunnelling points on the SC, on a
scale given by the superconducting coherence length $\xi$ which is
the size of a Cooper pair. This does not pose a severe restriction
for conventional s-wave materials  with $\xi$ typically being on
the order of $\mu {\rm m}$. In the relevant case $\delta r<\xi$
the suppression is only polynomial $\propto 1/(k_{F}\delta r)^2$,
with $k_{F}$ being the Fermi wave vector in the SC. On the other
hand, we see that the effect of the QDs  consists in the
suppression factor $(\gamma/{\cal E})^2$ for tunnelling into the
same lead \cite{cost}. Thus, in addition to Eq.~(\ref{regime}) we
have to impose the condition $k_F\delta r < {\cal E}/\gamma$,
which can be satisfied for small dots with ${\cal E}/\gamma\sim
100$ and
  $k_F^{-1}\sim 1\, {\rm \AA}$. As an experimental probe to test if
the two spin-entangled electrons indeed separate and tunnel to
different leads we suggest to join the two leads 1 and 2 to form
an Aharonov-Bohm loop. In such a setup the different tunnelling
paths of an Andreev process from the SC via the dots to the leads
can interfere. As a result, the measured current as a function of
the applied magnetic flux $\phi$ threading the loop contains a
phase coherent part $I_{AB}$ which consists of oscillations with
periods $h/e$ and $h/2e$ \cite{P3RSL}
\begin{equation}
\label{AB-osc} I_{AB}\sim
\sqrt{8I_{1}I_{2}}\cos(\phi/\phi_{0})+I_{2}\cos(2\phi/\phi_{0}),
\end{equation}
with $\phi_{0}=h/e$ being the single-electron flux quantum. The
ratio of the two contributions scales like $\sqrt{I_{1}/I_{2}}$
which suggest that by decreasing $I_{2}$ (e.g. by increasing $U$)
the $h/2e$ oscillations should vanish faster than the $h/e$ ones.

We note that the efficiency as well as the absolute rate for the
desired injection of two electrons into different  leads can even
be enhanced by using lower dimensional SCs \cite{P8RL,P9JS} . In
two dimensions (2D) we find that $I_{1}\propto 1/k_{F}\delta r$
for large $k_{F}\delta r$, and in one dimension (1D) there is no
suppression of the current and only an oscillatory behavior in
$k_{F}\delta r $ is found. A 2D-SC can be realized by using a SC
on top of a two-dimensional electron gas (2DEG) \cite{P4Klapwijk},
where superconducting correlations are induced via the proximity
effect in the 2DEG. In 1D, superconductivity was found in ropes of
single-walled carbon nanotubes \cite{P5Bouchiat}.

Finally, we note that the coherent injection of Cooper pairs by an
Andreev process allows the detection of individual spin-entangled
electron pairs in the leads. The delay time $\tau_{\rm delay}$
between the two electrons of a pair is given by $1/\Delta$,
whereas the separation in time of subsequent pairs is given
approximately by $\tau_{\rm pairs} \sim 2e/I_{1}\sim
\gamma_{l}/\gamma_{S}^{2}$ (up to geometrical factors)
\cite{P3RSL}. For $\gamma_{S}\sim\gamma_{l}/10\sim 1\mu {\rm eV}$
and $\Delta\sim 1{\rm meV}$ we obtain that the delay time
$\tau_{\rm delay}\sim 1/\Delta\sim 1{\rm ps}$ is much smaller than
the delivery time $\tau_{\rm pairs}$ per entangled pair
$2e/I_{1}\sim 40{\rm ns}$. Such a time separation is indeed
necessary in order to detect individual pairs of spin-entangled
electrons.

\subsubsection{Andreev entangler with Luttinger-liquid leads}
\label{sssluttinger}

\begin{figure}[h]
\centerline{\psfig{file=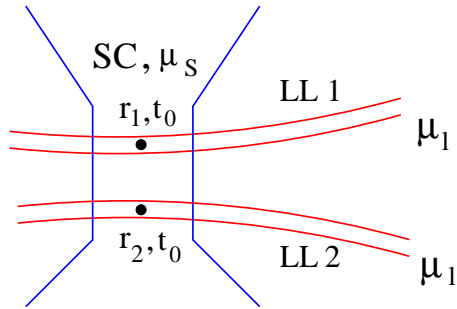,width=6cm}}
\vspace{5mm} \caption{Two quantum wires 1,2, with chemical
potential $\mu_{l}$ and described as infinitely long Luttinger
liquids (LLs), are deposited on top of an s-wave superconductor
(SC) with chemical potential $\mu_{S}$. The electrons of a Cooper
pair can tunnel by means of an Andreev process from two points
${\bf r}_{1}$ and ${\bf r}_{2}$ on the SC to the center (bulk) of
the two quantum wires 1 and 2, respectively, with tunnelling
amplitude $t_{0}$. Adapted from \cite{P8RL}.}\label{LLfig}
\end{figure}

Next we discuss a setup with an s-wave SC weakly coupled  to the
center (bulk) of two separate one-dimensional leads (quantum
wires) 1,2 (see Fig.~\ref{LLfig}) which exhibit Luttinger liquid
(LL) behavior, such as carbon nanotubes
\cite{P6Bockrath,P10Egger,P11Kane}. The leads are assumed to be
infinitely extended and are described by conventional LL-theory
\cite{P7rev}.

Interacting electrons in one dimension lack the existence of quasi
particles like they exist in a Fermi liquid and instead the low
energy excitations are collective charge and spin modes. In the
absence of backscattering interaction the velocities of the charge
and spin excitations are given by $u_{\rho}=v_{F}/K_{\rho}$ for
the charge and $u_{\sigma}=v_{F}$ for the spin, where $v_{F}$ is
the Fermi velocity and $K_{\rho}<1$ for repulsive interaction
between electrons ($K_{\rho}=1$ corresponds to a 1D-Fermi gas). As
a consequence of this non-Fermi liquid behavior, tunnelling into a
LL is strongly suppressed at low energies. Therefore one should
expect additional interaction effects in a two-particle tunnelling
event (Andreev process) of a Cooper pair from the SC to the leads.
We find that strong LL-correlations result in an additional
suppression for tunnelling of two coherent electrons into the {\em
same} LL compared to single electron tunnelling into a LL if the
applied bias voltage $\mu$ between the SC and the two leads is
much smaller than the energy gap $\Delta$ of the SC.

To quantify the effectiveness of such an entangler, we calculate
the current for the two competing processes of tunnelling into
different leads ($I_{1}$) and into the same lead ($I_{2}$) in
lowest order via a tunnelling Hamiltonian approach. Again we
account for a finite distance separation $\delta r$ between the
two exit points on the SC when the two electrons of a Cooper pair
tunnel to different leads. For the current $I_{1}$ of the desired
pair-split process we obtain, in leading order in $\mu/\Delta$ and
at zero temperature~\cite{P8RL,P9JS}
\begin{equation}
\label{LL1}
I_{1}=\frac{I_{1}^{0}}{\Gamma(2\gamma_{\rho}+2)}\frac{v_{F}}{u_{\rho}}
\left[\frac{2\Lambda\mu}{u_{\rho}}\right]^{2\gamma_{\rho}},
\,\,I_{1}^{0}=\pi e\gamma^{2}\mu F_{d}[\delta r],
\end{equation}
where $\Gamma (x)$ is the Gamma function and $\Lambda$ is a short
distance cut-off on the order of the lattice spacing in the LL and
$\gamma=4\pi\nu_{S}\nu_{l}|t_{0}|^{2}$ is the dimensionless tunnel
conductance per spin with $t_{0}$ being the bare tunnelling
amplitude for electrons to tunnel from the SC to the LL-leads (see
Fig.~\ref{LLfig}). The electron density of states per spin at the
Fermi level for the SC and the LL-leads are denoted by $\nu_{S}$
and $\nu_{l}$, respectively. The current $I_{1}$ has its
characteristic non-linear form $I_{1}\propto
\mu^{2\gamma_{\rho}+1}$ with
$\gamma_{\rho}=(K_{\rho}+K_{\rho}^{-1})/4-1/2>0$ being the
exponent for tunnelling into the bulk of a {\em single} LL. The
factor $F_{d}[\delta r]$ in (\ref{LL1}) depends on the geometry of
the device and is given here again by $F_{d}[\delta
r]=[\sin(k_{F}\delta r)/k_{F}\delta r]^{2}\exp(-2\delta r/\pi\xi)$
for the case of a 3D-SC. In complete analogy to
subsection~\ref{ssecAndreev} the power law suppression in
$k_{F}\delta r$ gets weaker in lower dimensions.

This result should be compared with the unwanted transport channel
where two electrons of a Cooper pair tunnel into the same lead 1
or 2 but with $\delta r=0$.  We find that such processes are
indeed suppressed by strong LL-correlations if $\mu<\Delta$. The
result for the current ratio $I_{2}/I_{1}$ in leading order in
$\mu/\Delta$ and for zero temperature is \cite{P8RL,P9JS}
\begin{equation}
\frac{I_{2}}{I_{1}}=F_{d}^{-1}[\delta r]\sum\limits_{b=\pm
1}\,A_{b}\,\left(\frac{2\mu}{\Delta}\right)^{2\gamma_{\rho
b}},\,\,\gamma_{\rho +}=\gamma_{\rho},\,\,\gamma_{\rho
-}=\gamma_{\rho}+(1-K_{\rho})/2, \label{currentI222}
\end{equation}
where $A_{b}$ is an interaction dependent
constant~\cite{LLfootnote}. The result (\ref{currentI222}) shows
that the current $I_{2}$ for injection of two electrons into the
same lead is suppressed compared to $I_{1}$ by a factor of
$(2\mu/\Delta)^{2\gamma_{\rho +}}$, if both electrons are injected
into the  same branch (left or right movers), or by
$(2\mu/\Delta)^{2\gamma_{\rho -}}$ if the two electrons travel in
different directions \cite{P21electronbunching}. The suppression
of the current $I_{2}$ by $1/\Delta$ reflects the two-particle
correlation effect in the LL, when the electrons tunnel into the
same lead. The larger $\Delta$, the shorter the delay time is
between the arrivals of the two partner electrons of a Cooper
pair, and, in turn, the more the second electron tunnelling into
the same lead will feel the existence of the first one which is
already present in the LL. This behavior is similar to the Coulomb
blockade effect in QDs, see subsection~\ref{ssecAndreev}. Concrete
realizations of LL-behavior is found in metallic carbon nanotubes
with similar exponents as derived here \cite{P10Egger,P11Kane}. In
metallic single-walled carbon nanotubes $K_{\rho}\sim 0.2$
\cite{P6Bockrath} which corresponds to $2\gamma_{\rho}\sim  1.6$.
This suggests the rough estimate $(2\mu/\Delta)<1/k_{F}\delta r$
for the entangler to be efficient. As a consequence, voltages in
the range $k_{B}T<\mu<100 \mu {\rm eV}$ are required for $\delta
r\sim$ nm and $\Delta\sim 1{\rm meV}$.  In addition, nanotubes
were reported to be very good spin conductors \cite{P12Balents}
with estimated spin-flip scattering lengths of the order of $\mu
{\rm m}$ \cite{P13BVBF}.

We remark that in order to use the beam-splitter setup to detect
spin-entanglement via noise the two LL-leads can be coupled
further to Fermi liquid leads. In such a setup the LL-leads then
would act as QDs \cite{spinexc}. Another way to prove
spin-entanglement is to carry out spin-dependent current-current
correlation measurements between the two LLs. Such spin dependent
currents can be measured e.g. via spin filters (Sec.
\ref{spinfilters}).

\subsection{Triple-quantum dot entangler}
\label{triple-dot-entangler}

\begin{figure}[!b]
\psfig{file=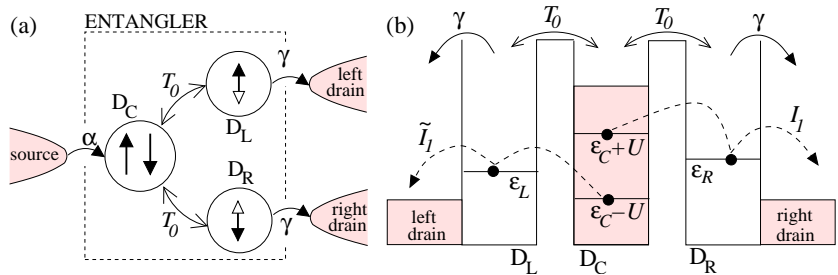,width=11cm} \caption{{\footnotesize (a)
Setup of the triple quantum dot entangler. Three leads are coupled
to three quantum dots in the Coulomb blockade regime. The central
dot \dc can accept \( 0 \), \(1 \) or \( 2 \) electrons provided
from the source lead with rate \( \alpha \); with 2 electrons, its
ground state is the spin singlet. The tunnelling amplitudes \(
T_0\) describe the coherent tunnelling between \dc and the
secondary dots \dl and \dr, which can only accept \( 0 \) or \( 1
\) electron. Each electron can finally tunnel out to the drain
leads with a rate \( \gamma \). (b)} \emph{\footnotesize
Single-particle} {\footnotesize energy level diagram. The dashed
arrows represent the single-electron currents \( I_1 \) and \(
\tilde{I_1} \). Adapted from \cite{Sar02}. \vspace{-0mm}}}
\label{figset}
\end{figure}

In this proposal \cite{Sar02}, the pair of spin-entangled
electrons is provided by the ground state of a single quantum dot
\dc with an even number of electrons, which is the spin-singlet
\cite{singlet}; see Fig. \ref{figset}.
In the Coulomb blockade regime \cite{Kou97}, electron interactions
in each dot create a large charging energy \(U\) that provides the
energy filtering necessary for the suppression of the
non-entangled currents. These arise either from the escape of the
pair to the same lead, or from the transport of a single electron.
The idea is to create a resonance for the joint transport of the
two electrons from \dc to secondary quantum dots \dl and \(D_R\),
similarly to the resonance described in Sec. \ref{ssecAndreev} .
For this, we need the condition \( \el+\er=2 \ec \), where \( \el
\) and \( \er \) are the energy levels of the available state in
\dl and \(D_R\), and \(2 \ec \) is the total energy of the two
electrons in \(D_C\). On the other hand, the transport of a single
electron from \dc to \dl or \dr is suppressed by the energy
mismatch \( \ec\pm U \neq \el, \er \), where \( \ec\pm U \) is the
energy of the \( 2^{\rm nd} / 1^{\rm st} \) electron in
\dc\cite{charging}.

\begin{figure}[!htb]
\begin{center} \psfig{file=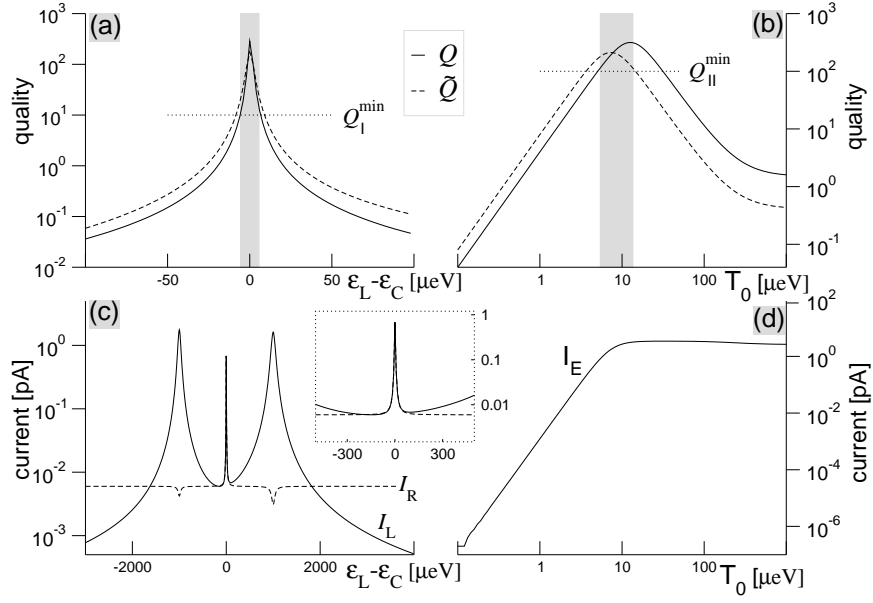,width=9cm,angle=270} \end{center} \caption{{\small
Quality and current of the entangler, with the parameters \(
\alpha =0.1,\gamma =1,T_{0}=10,U=1000 \) in \( \mu \mathrm{eV} \).
(a) Quality \( Q \) and \( \widetilde{Q} \), around the resonance
at \( \el-\ec=0 \) where the entangled current dominates. In gray,
the width of the resonance defined by \(
Q,\widetilde{Q}>Q_{\mathrm{I}}^{\mathrm{min}}=10 \) is \(
|\el-\ec|< 6\, \mu \mathrm{eV} \), as predicted by
Eq.(\ref{qual1}). (b) \( Q \) and \( \widetilde{Q} \) as a
function of \( T_{0} \) at resonance (\( \el=\ec \)). In gray, the
region where the quality of the entangler is \(
Q,\widetilde{Q}>Q_{\mathrm{II}}^{\mathrm{min}}=100 \)
corresponding to Eq. (\ref{qual2}). (c) Entangled and
non-entangled current in the left (\( I_{L} \)) and in the right
(\( I_{R} \)) drain leads. The inset shows the resonance in a
larger scale. (d) Saturation of the entangled current \( I_{E} \).
Adapted from \cite{Sar02}. \vspace{-0mm}}} \label{figres}
\end{figure}

We describe the incoherent sequential tunneling between the leads
and the dots in terms of a master equation \cite{Blu96} for the
density matrix \( \rho\) of the triple-dot system (valid for
$k_{\rm B}T
> \gamma$). The stationary solution of the master equation is found with
MAPLE, and is used to define stationary currents. Besides the
entangled current \(I_E\) coming from the {\em joint} transport of
the electrons from \dc to \dl and \(D_R\), the solitary escape of
one electron of the singlet can create a non-entangled current \(
I_1 \), as it could allow a new electron coming from the source
lead to form a new spin-singlet with the remaining electron.
Another non-entangled current \( \tilde{I}_1 \) can be present if
only one electron is transported across the triple-dot system; see
Fig. \ref{figset}(b). The definition of entangler {\em qualities}
\( Q=I_E/I_1 \) and \( \tilde{Q}=I_E/\tilde{I_1} \) enables us to
check the suppression of these non-entangled currents.

In Fig. \ref{figres} we present results in the case where \(
\er=\ec \). This gives a two-electron resonance at \( \el=\ec=\er
\), and create mobile entangled electrons with the same orbital
energy, as required in the beam-splitter setup to allow
entanglement detection \cite{gde}, \cite{prl-cgd}. The exact
analytical expressions are extremely lengthy, but we can get
precise conditions for an efficient entangler regime by performing
a Taylor expansion in terms of \( \alpha ,\gamma ,T_{0} \)
(defined in Fig. \ref{figset}). Introducing the conditions \(
Q,\widetilde{Q}>Q^{\mathrm{min}}_{\mathrm{I}} \) away from
resonance (\( \el \neq \ec \)) and \(
Q,\widetilde{Q}>Q^{\mathrm{min}}_{\mathrm{II}} \) at resonance (\(
\el = \ec \)), we obtain the conditions \cite{Sar02}
\begin{eqnarray}
&& | \el-\ec |< 2T_{0}/\sqrt{Q^{\mathrm{min}}_{\mathrm{I}}} \ ,
\label{qual1}\\ &&\gamma
\sqrt{Q^{\mathrm{min}}_{\mathrm{II}}/8}<T_{0}<U\sqrt{4\alpha /
\gamma Q_{\mathrm{II}}^{\mathrm{min}}}. \label{qual2}
\end{eqnarray}
We need a large \( U \) for the energy suppression of the
one-electron transport, and
 \( \gamma \ll T_0 \) because  the joint transport is a higher-order process in \(T_0\).
The current saturates to \( I_E \to e \alpha \) when \(
T^{4}_{0}\gg U^{2}\gamma \alpha /32 \) [see Fig. \ref{figres}(d)]
when the bottleneck process is the tunneling of the electrons from
the source lead to the central dot. We see in (c) that equal
currents in the left and right drain lead, \( I_{L} = I_R\), are
characteristic of the resonance \( \el= \ec \), which provides an
experimental procedure to locate the efficient regime.

Taking realistic parameters for quantum dots \cite{Kou97,Exp} such
as \( I_E = 20 \) pA, \( \alpha =0.1\) $\mu$eV and \( U=1 \) meV,
 we obtain
a maximum entangler quality \( Q_{\mathrm{II}}^{\mathrm{min}}=100
\) at resonance, and a finite width \( | \el-\ec |\simeq 6\)
$\mu$eV where the quality is at least \(
Q_{\mathrm{I}}^{\mathrm{min}}=10 \). Note that one must avoid
resonances with excited levels which could favour the undesired
non-entangled one-electron transport. For this, one can either
tune the excited levels away by applying a magnetic field, or
require a large energy levels spacing \( \Delta \epsilon _{i}
\simeq 2 U\), which can be found in vertical quantum dots or
carbon nanotubes \cite{Kou97}. We can estimate the relevant
timescales by simple arguments. The entangled pairs are delivered
every \( \tau_{\mathrm{pairs}}\simeq 2/\alpha \simeq 13\,
\mathrm{ns} \). The average separation between two entangled
electrons within one pair is given by the time-energy uncertainty
relation: \( \tau_{\mathrm{delay}}\simeq 1/U\simeq 0.6\,
\mathrm{ps} \), while their maximal separation is given by the
variance of the exponential decay law of the escape into the
leads:
 \( \tau_{\mathrm{max}}\simeq 1/\gamma \simeq 0.6\, \mathrm{ns} \).
Note that \( \tau_{\mathrm{delay}} \) and \(  \tau_{\mathrm{max}}
\) are both well below reported spin decoherence times (in bulk)
of \( 100\, \mathrm{ns} \) \cite{Kik97}.

\section{Spin-polarized electron sources}
\label{spinfilters}

Here we briefly mention some of the possibilities for
spin-polarized electron sources possibly relevant as feeding
Fermi-liquid reservoirs to our beam-splitter configuration. Even
though we are concerned here with mesoscopic \textit{coherent}
transport, we emphasize that the electron sources themselves can
be diffusive or ballistic.

Currently, there is a great deal of interest in the problem of
spin injection in hybrid mesoscopic structures. At the simplest
level we can say that the ``Holy Grail'' here is essentially the
ability to spin inject \textit{and} detect spin-polarized charge
flow across interfaces. The possibility of controlling and
manipulating the degree of spin polarization of the flow is highly
desirable. This would enable novel spintronic devices with
flexible/controllable functionalities.

Recently, many different experimental possibilities for spin
injection/detection have been considered: (i) all-optical
\cite{irina} and (ii) all-electrical \cite{johnsson},
\cite{jedema} spin injection and detection in semiconductors and
metal devices, respectively, and (iii) electric injection with
optical detection in hybrid (Mn-based) ferromagnetic/non-magnetic
and paramagnetic/non-magnetic semiconductor \textit{pin} diodes
\cite{ohno-molen}. For an account of the experimental efforts
currently underway in the field of spin-polarized transport, we
refer the reader to Ref. \cite{als-review}. Below we focus on our
proposals for spin filtering with a semimagnetic tunnel barrier
\cite{egues} and a quantum dot \cite{patrik}. These can, in
principle, provide alternative schemes for spin injection into our
beam splitter.

\subsection{Quantum spin filtering}

Ballistic Mn-based tunnel junctions \cite{egues} offer an
interesting possibility for generating spin-polarized currents.
Here the s-d interaction in the paramagnetic layer gives rise to a
spin-dependent potential. An optimal design can yield high
barriers for spin-up and vanishingly small barriers for spin-down
electrons. Hence, a highly spin-selective tunnel barrier can be
achieved in the presence of an external magnetic field. Note that
here \textit{ballistic spin filtering} -- due to the blocking of
one spin component of the electron flow -- is the relevant
mechanism for producing a spin-polarized current. Earlier
calculations have shown that full spin polarizations are
attainable in ZnSe/ZnMnSe spin filters \cite{egues}.

\subsection{Quantum dots as spin filters}

Spin polarized currents can also be generated by a quantum dot
\cite{patrik}. In the Coulomb blockade regime with Fermi-liquid
leads, it can be operated as an efficient spin-filter
\cite{patrik-sf-memory} at the single electron level. A magnetic
field lifts the spin degeneracy in the dot while its effect is
negligible \cite{patrik-sf-filter} in the leads. As a consequence,
only one spin direction can pass through the quantum dot from the
source to the drain. The transport of the opposite spin is
suppressed by energy conservation and singlet-triplet splitting.
This filtering effect can be enhanced by using materials with
different g-factors for the dot and the lead. To increase the
current signal, one could also use an array of quantum dots, e.g.
self-assembled dots.

\subsection{spin filters for spin detection and Bell inequalities}
\label{bell}

Besides being a source of spin-polarized currents, such spin
filters (with or without spin-polarized sources
\cite{patrik},\cite{hanres}) could be used to measure electron
spin, as they convert spin information into charge: the
transmitted charge current depends on the spin direction of the
incoming electrons \cite{ld}. Such filters could probe the degree
of polarization of the incoming leads. In addition, Bell
inequalities measurements could be performed with such devices
\cite{P14Kawabata,P15Martin}.

\section{Scattering formalism: basics}

\textit{Current.} In a multi-probe configuration with incoming and
outgoing leads related via the scattering matrix
$\mathbf{s}_{\gamma \beta } $, the current operator in lead
$\gamma $ within the Landauer-B\"uttiker \cite{buttiker} approach
is given by
\begin{eqnarray}
\hat{I}_{\gamma }(t) &=&\frac{e}{h}\sum_{\alpha \beta }\!\!\int
\!\!d\varepsilon d\varepsilon ^{\prime}e^{i(\varepsilon
-\varepsilon^{\prime})t/\hbar }\mathbf{a}_{\alpha }^{\dagger }(\varepsilon )\mathbf{A}%
_{\alpha ,\beta }(\gamma ;\varepsilon ,\varepsilon ^{\prime})
\mathbf{a}_{\beta }(\varepsilon ^{\prime}), \nonumber \\
&&\mathbf{A}_{\alpha \beta }(\gamma ;\varepsilon
,\varepsilon^{\prime})=\delta _{\gamma \alpha }\delta _{\gamma
\beta }\mathbf{1}-\mathbf{s}_{\gamma \alpha }^{\dagger
}(\varepsilon )\mathbf{s}_{\gamma \beta }(\varepsilon ^{\prime }),
\label{ceq1}
\end{eqnarray}
where we have defined the two-component object $
\mathbf{a}^\dagger_{\alpha } (\varepsilon)=(a^\dagger_{\alpha,
\uparrow} (\varepsilon), a^\dagger_{\alpha, \downarrow}
(\varepsilon))$ with $ a^\dagger_{\alpha, \sigma} (\varepsilon)$
denoting the usual fermionic creation operator for an electron
with energy $ \varepsilon $ and spin component
$\sigma=\uparrow,\downarrow $ in lead $ \alpha $. Here the spin
components $\sigma $ are along a properly defined quantization
axis (e.g., $x$, $y$ or $z$).

\textit{Noise.} Let $\delta \hat{I}_{\gamma }(t)= \hat{I}_{\gamma
}(t) - \langle I \rangle$ denote the current-fluctuation operator
at time $t$ in lead $\gamma$ ($\langle I \rangle$: average
current). We define noise between leads $\gamma$ and $\mu$ in a
multi-terminal system by the average power spectral density of the
symmetrized current-fluctuation autocorrelation function
\cite{factor-of-two}
\begin{equation}
S_{\gamma \mu }(\omega )=\frac{1}{2}\int \langle \delta
\hat{I}_{\gamma }(t)\delta \hat{I}_{\mu }(t^{\prime })+\delta
\hat{I}_{\mu }(t^{\prime })\delta \hat{I}_{\gamma }(t)\rangle
e^{i\omega t}dt.  \label{ceq2}
\end{equation}
The angle brackets in Eq. (\ref{ceq2}) denote either an ensemble
average or an expectation value between relevant pairwise electron
states. We focus on noise at zero temperatures. In this regime,
the current noise is solely due to the discreteness of the
electron charge and is termed \textit{shot noise}.

\subsection{Scattering matrix}

\textit{Electron beam splitter.} This device consists of four
quasi one-dimensional leads (point contacts) electrostatically
defined on top of a 2DEG \cite{bs}, \cite{henny}. An extra
``finger gate'' in the central part of the device acts as a
potential barrier for electrons traversing the system, i.e., a
``beam splitter''. That is, an impinging electron from, say, lead
1 has probability amplitudes $r$ to be reflected into lead 3 and
$t$ to be transmitted into lead 4.

\textit{Beam splitter $\mathbf{s}$ matrix.} The transmission
processes at the beam splitter can be suitably described in the
language of the scattering theory: $s_{13}=s_{31}=r$ and
$s_{14}=s_{41}=t$; similarly, $s_{23}=s_{32}=t$ and
$s_{24}=s_{42}=r$, see Fig. \ref{fig2c}. We also neglect
backscattering into the incoming leads,
$s_{12}=s_{34}=s_{\alpha\alpha}=0$.
Note that the beam splitter $\mathbf{s}$ matrix is spin
independent; this no longer holds in the presence of a spin-orbit
interaction. We also assume that the amplitudes $r$ and $t$ are
energy independent. The unitarity of $\mathbf{s}$ implies
$|r|^2+|t|^2=1$ and $\mathrm{Re}(r^\ast t)=0$. Below we use the
above scattering matrix to evaluate noise.

\section{Noise of entangled electron pairs: earlier results}
\label{earlier-results}

\begin{figure}[h]
\begin{center}
\epsfig{file=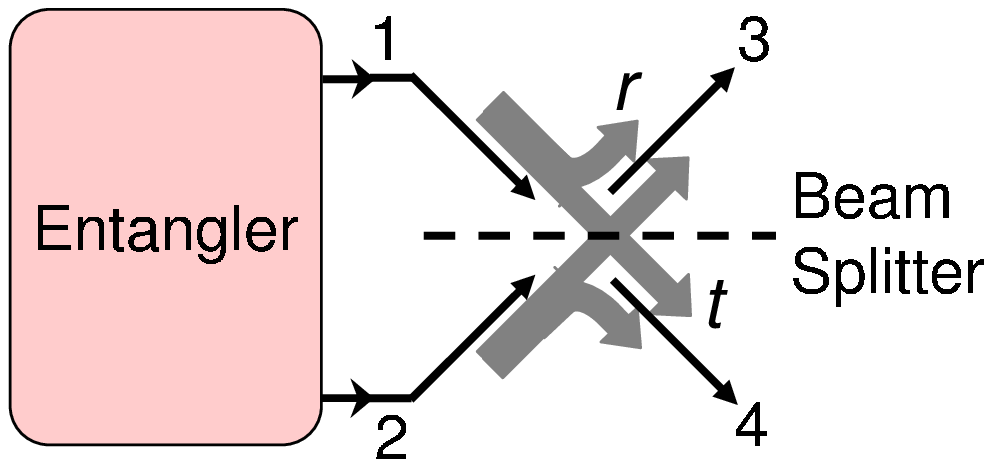,width=0.95\textwidth}
\end{center}
\par
\vspace{-5mm} \caption{Electron entangler coupled to a beam
splitter device. ``Entangler'' here represents one of the proposed
setups of Sec. \ref{entanglers} or some other arrangement
providing both triplet and singlet pairs via proper level tuning
with gate electrodes. Adapted from Ref. \cite{gde}.} \label{fig2c}
\end{figure}

\textit{Singlet and triplets.} Let us assume that an entangler is
now ``coupled'' to the beam-splitter device so as to inject
entangled (and unentangled) electron pairs into the incoming
leads, Fig. \ref{fig2c}. This will certainly require some
challenging lithographic patterning and/or elaborate gating
structures.

Let us consider the following two-electron states
\begin{equation}
|S\rangle =\frac{1}{\sqrt{2}}\left[ a_{1\uparrow }^{\dagger
}(\varepsilon _{1})a_{2\downarrow }^{\dagger }(\varepsilon
_{2})-a_{1\downarrow }^{\dagger }(\varepsilon _{1})a_{2\uparrow
}^{\dagger }(\varepsilon _{2})\right] |0\rangle ,\label{ceq4}
\end{equation}
\begin{equation}
|Te\rangle =\frac{1}{\sqrt{2}}\left[ a_{1\uparrow }^{\dagger
}(\varepsilon _{1})a_{2\downarrow }^{\dagger }(\varepsilon
_{2})+a_{1\downarrow }^{\dagger }(\varepsilon _{1})a_{2\uparrow
}^{\dagger }(\varepsilon _{2})\right] |0\rangle ,  \label{ceq5}
\end{equation}
and
\begin{equation}
|Tu_{\sigma }\rangle =a_{1\sigma }^{\dagger }(\varepsilon
_{1})a_{2\sigma }^{\dagger }(\varepsilon _{2})|0\rangle , \qquad
\sigma =\uparrow , \downarrow .  \label{ceq6}
\end{equation}
The above states correspond to the singlet $|S\rangle$, the
entangled triplet $|Te\rangle $, and the unentangled triplets
$|Tu_{\sigma }\rangle$, respectively, injected electron pairs.
Note that $|0\rangle$ denotes the ``lead vacuum'', i.e., an empty
lead or a Fermi sea. Here we follow Ref. \cite{gde} and assume
that the injected pairs have \textit{discrete} energies
$\varepsilon_{1,2}$.

To determine the average current and shot noise for electron pairs
we have to calculate the expectation value of the noise
two-particle states in Eqs. (\ref{ceq4})-(\ref{ceq6}). In the
limit of zero bias, zero temperature, and zero frequency, we find
\cite{gde}
\begin{equation}
S_{33}^{S/Te,u_{\sigma }}=\frac{2e^{2}}{h\nu }T(1-T)(1\pm \delta
_{\varepsilon _{1},\varepsilon _{2}}),  \label{ceq7}
\end{equation}
for the shot noise in lead 3 for singlet (upper sign) and triplets
(lower sign) with $T\equiv |t|^2$ (transmission coefficient). The
corresponding currents in lead 3 are
$I_3^{S,Te,u_{\sigma }}=I=\frac{e}{h\nu }$.
Note the density of states factor $\nu$ in Eqs. (\ref{ceq7})
arising from the discrete spectrum used \cite{discrete}.

\textit{Bunching and antibunching.} For $\varepsilon
_{1}=\varepsilon _{2}$ the Fano factors corresponding to the shot
noise in Eq. (\ref{ceq7}) are
$F^{S}=S_{33}^S/eI=4T(1-T)$,
for the singlet and
$F^{Te,u_{\sigma }}=0$,
for all three triplets. Interestingly, the Fano factor for a
singlet pair is enhanced by a factor of two as compared to the
Fano factor $2T(1-T)$ for a \textit{single} uncorrelated electron
beam \cite{fu} impinging on the beam splitter; the Fano factor for
the triplets is suppressed with respect to this uncorrelated case.
This enhancement of $F^S$ and suppression of $F^{Te,u_{\sigma }}$
is due to \textit{bunching} and \textit{antibunching},
respectively, of electrons in the outgoing leads. This result
offers the possibility of distinguishing singlet from triplet
states via noise measurements (triplets cannot be distinguished
among themselves here; a further ingredient is needed for this,
e.g., a local Rashba interaction in one of the incoming leads).

\section{Electron transport in the presence of a \textit{local} Rashba s-o interaction}

The central idea here is to use the gate-controlled Rashba
coupling to rotate the electron spins \cite{datta-das} traversing
the Rashba-active region (lead 1 of the beam splitter), thus
altering in a controllable way the resulting transport properties
of the system. Below we first discuss the effects of the Rashba
s-o interaction in one-dimensional systems; the incoming leads are
essentially quasi one-dimensional wires, i.e, ``quantum point
contacts''. A local Rashba interaction can in principle be
realized with an additional gating structure (top and back gates
\cite{gatecontrol}).

We focus on wires with one and two transverse channels
\cite{moroz}. This latter case allows us to study the effects of
s-o induced interband coupling on both current and shot noise.

\subsection{Rashba wires with uncoupled transverse channels}

\subsubsection{Hamiltonian, eigenenergies and eigenvectors}

The Rashba spin-orbit interaction is present in low-dimensional
systems with \textit{structural} inversion asymmetry. Roughly
speaking, this interaction arises from the gradient of the
confining potential (``triangular shape'') at the interface
between two different materials \cite{electric-field}. For a
non-interacting one-dimensional wire with \textit{uncoupled}
transverse channels, the electron Hamiltonian in the presence of
the Rashba coupling $\alpha$ reads \cite{rashba}
\begin{equation}
H_n=-\frac{\hbar^2}{2m^{\ast}}\partial^2_x+ \epsilon_n +
i\alpha\sigma_y
\partial_x. \label{ceq11}
\end{equation}
In Eq. (\ref{ceq11}) $\partial_x\equiv\partial/\partial x$,
$\sigma_y$ is the Pauli matrix, $m^{\ast}$ is the electron
effective mass, and $\epsilon_n$ is the bottom of the
n$^{th}$-channel energy band in absence of s-o interaction. For an
infinite-barrier transverse confinement of width $w$,
$\epsilon_n=n^2\pi^2 \hbar^2/(2mw^2)$.

\begin{figure}[h]
\begin{center}
\epsfig{file=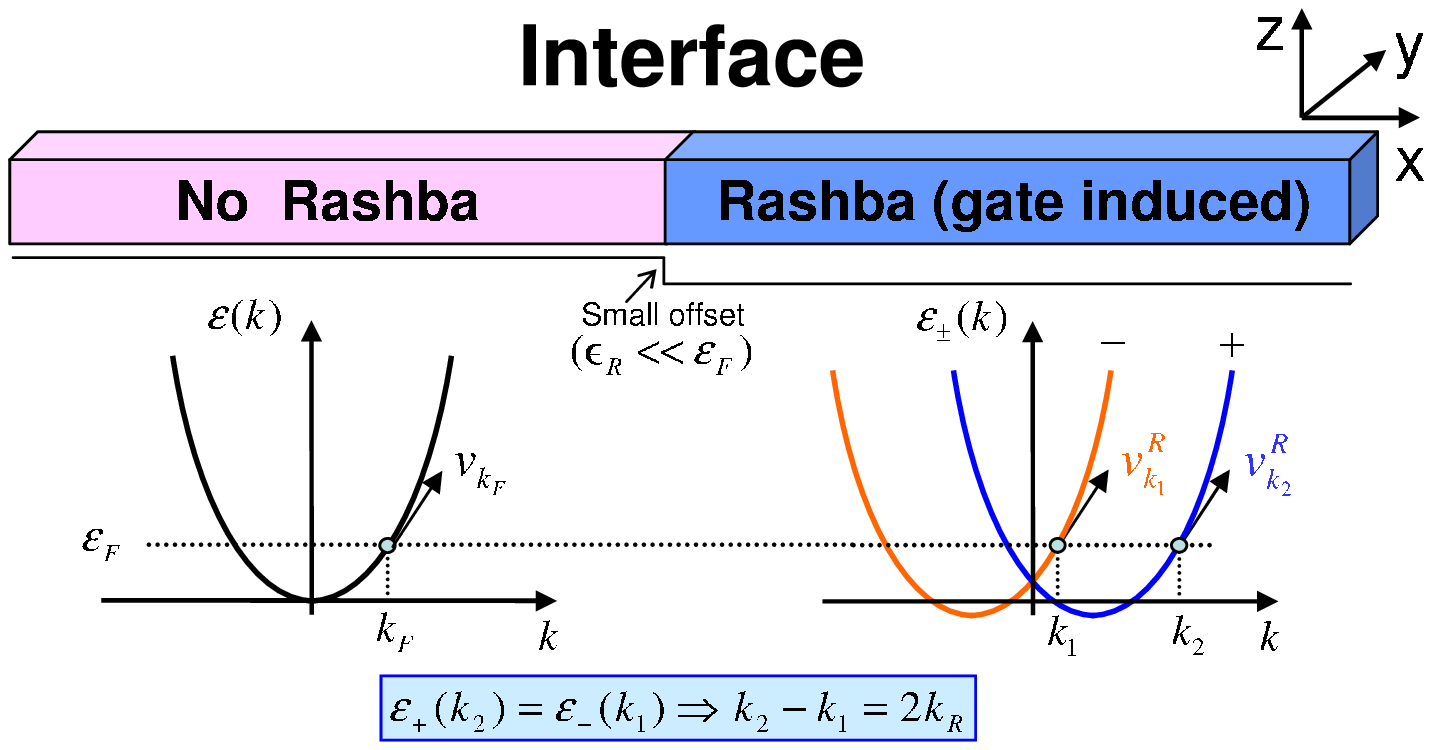,width=0.95\textwidth}
\end{center}
\par
\vspace{-5mm} \caption{Schematic of a portion of a gate-induced
 no-Rashba/Rashba ``interface'' and its corresponding band
structure. Note the small band offset arising solely from the
mismatch $\epsilon_R$.} \label{fig3c}
\end{figure}

The Hamiltonian in (\ref{ceq11}) yields the usual set of Rashba
bands \cite{molen-flesh}
\begin{equation}
\varepsilon^n_s =\hbar^{2}(k - sk_{R})^{2}/2m^{\ast} + \epsilon_n
-\epsilon_{R}, \qquad s=\pm \label{ceq12}
\end{equation}
where $k_{R}=m^{\ast}\alpha/\hbar^2$ and $\epsilon _{R}=\hbar
^{2}k_{R}^{2}/2m^{\ast }=m^{\ast}\alpha^2/2\hbar^2$ (``Rashba
energy''). The corresponding wave functions are eigenvectors of
$\sigma_y$ with the orbital part being a plane wave times the
transverse-channel wave function. Figure \ref{fig3c} shows that
the parabolic bands are shifted sideways due to the Rashba
interaction. Note that these bands are still identified by a
unique spin index $s=\pm$ which in our convention corresponds to
the eigenspinors  $|\mp \rangle \sim |\uparrow\rangle \mp |
\downarrow \rangle$ of $\sigma_y$.

\subsubsection{Boundary conditions and spin injection}
\label{one-band-injection}

Here we assume a unity transmission across the interface
\cite{unity} depicted in Fig. \ref{fig3c}. For a spin-up electron
with wave vector $k_F$ entering the Rashba region at $x=0$, we
have the following boundary conditions for the wave function and
its derivative \cite{molen-flesh,bc}
\begin{equation}
|\uparrow \rangle e^{ik_Fx}|_{x\rightarrow 0^-} =
\frac{1}{\sqrt{2}}[|+ \rangle e^{ik_2x}+|- \rangle
e^{ik_1x}]_{x\rightarrow 0^+}, \label{ceq13}
\end{equation}
and
\begin{equation}
|\uparrow \rangle v_{k_F} e^{ik_Fx}|_{x\rightarrow 0^-} =
\frac{1}{\sqrt{2}}[|+ \rangle v^R_{k_2}e^{ik_2x}+|- \rangle
v^R_{k_1}e^{ik_1x}]_{x\rightarrow 0^+}, \label{ceq14}
\end{equation}
with the Fermi and Rashba group velocities defined by
$v_{k_F}=\hbar k_F/m^{\ast}$,
$v^R_{k_1}=\frac{\hbar}{m^{\ast}}(k_1+k_R)$, and
$v^R_{k_2}=\frac{\hbar}{m^{\ast}}(k_2-k_R)$. The wave vectors
$k_1$ and $k_2$ are defined by the ``horizontal'' intersections
with the Rashba bands $\varepsilon_{-}(k_1)=
\varepsilon_{+}(k_2)$, see Fig. \ref{fig3c}. This results in the
condition $k_2-k_1=2k_R$ which implies that the Rashba group
velocities are the same at these points: $v^R_{k_1}=v^R_{k_2}$.
Equation (\ref{ceq14}) is satisfied provided that
\cite{molen-flesh}
\begin{equation}
v_{k_F}=\frac{1}{2}(v^R_{k_1}+v^R_{k_2})=\frac{\hbar}{m^{\ast}}
\sqrt{\frac{2m^{\ast}}{\hbar^2}(\varepsilon_F+\epsilon_R)},
\label{ceq18}
\end{equation}
where the last equality follows from conservation of energy,
$\varepsilon_{-}(k_1)= \varepsilon_{+}(k_2)=\varepsilon_F$. Note
that the group velocity of the incoming spin-up electron is
completely ``transferred'' to the Rashba states at the interface.

\textit{Spin-rotated state at $x=L$.} For an incoming spin-up
electron, we have at the exit of the Rashba region the
spin-rotated state
\begin{equation}
\psi_{\uparrow,L} =\frac{1}{\sqrt{2}}[|+ \rangle e^{ik_2L}+|-
\rangle e^{ik_1L}], \label{ceq19}
\end{equation}
which is consistent with the boundary conditions (\ref{ceq13}) and
(\ref{ceq14}). After some straightforward manipulations (and using
$k_2-k_1=2k_R$), we find
\begin{equation}
\psi_{\uparrow,L}  =\left(
\begin{array}{c}
\cos \theta _{R}/2 \\
\sin \theta _{R}/2
\end{array}
\right) e^{i(k_1+k_R)L},  \label{ceq20}
\end{equation}
with the usual Rashba angle $\theta_R=2m^{\ast}\alpha L/\hbar^2$
\cite{datta-das,apl-cgd}. A similar expression holds for an
incoming spin-down electron. Note that the boundary conditions at
$x=L$ are trivially satisfied since we assume unity transmission.
The overall phase of the spinor in Eq. (\ref{ceq20}) is irrelevant
for our purposes; we shall drop it from now on.

\subsubsection{Rashba spin rotator}

From the results of the previous section we can now define a
unitary operator which describes the action of the Rashba-active
region on any incoming spinor
\begin{equation}
\mathbf{U_{R}}=\left(
\begin{array}{cc}
\cos \theta _{R}/2 & -\sin \theta _{R}/2 \\
\sin \theta _{R}/2 & \cos \theta _{R}/2%
\end{array}
\right).  \label{ceq21}
\end{equation}
Note that all uncoupled transverse channels are described by the
same unitary operator $\mathbf{U_{R}}$. The above unitary operator
allows us to incorporate the s-o induced precession effect
straightforwardly into the scattering formalism (Sec.
\ref{beam-splitter-rashba}).

\subsection{Rashba wire with two coupled transverse channels}
\label{two-bands}

The Rashba s-o interaction also induces a coupling between the
bands described in the previous section. Here we extend our
analysis to the case of two \textit{weakly} coupled Rashba bands.

\subsubsection{Exact and approximate energy bands}

\begin{figure}[h]
\begin{center}
\epsfig{file=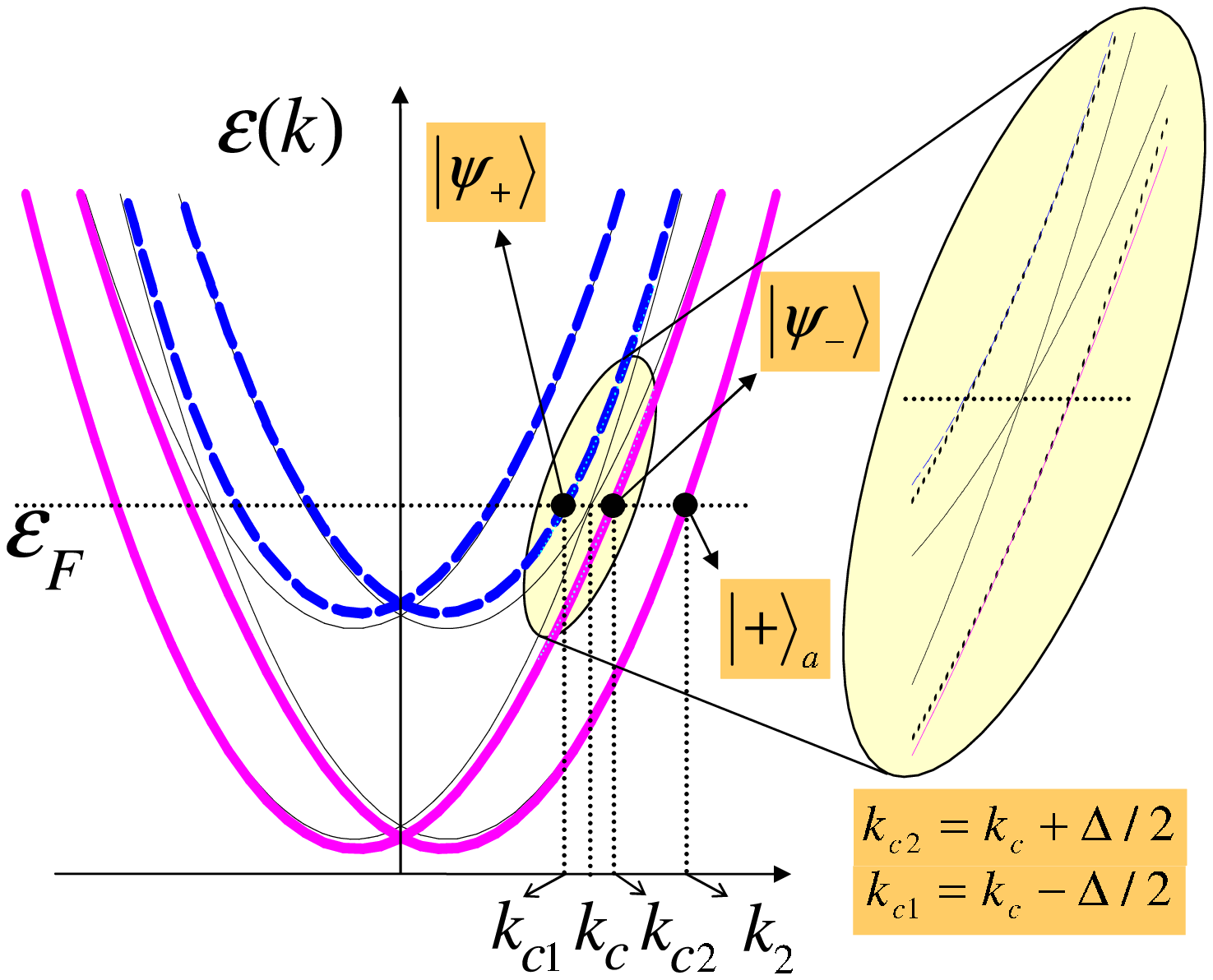,width=0.8\textwidth}
\end{center}
\par
\vspace{-5mm} \caption{Band structure for a wire with two sets of
Rashba bands. Both the uncoupled (thin lines) and the
interband-coupled (thick solid and dashed lines) are shown. The
uncoupled Rashba bands cross at $k_c$. Spin-orbit induced
interband coupling gives rise to anti crossing of the bands near
$k_c$. Inset: blowup of the region near the crossing. The
nearly-free electron bands [perturbative approach, Eq.
(\ref{ceq24})] describe quite well the exact dispersions near the
crossing (cf. dotted and solid + dashed lines in the inset). The
solid circles (``intersections'') indicate the relevant $k$ points
for spin injection [Eq. (\ref{ceq26a})]; their corresponding
zeroth-order eigenvectors [Eq. (\ref{ceq25})] are also indicated.}
\label{fig5c}
\end{figure}

Projecting the two-dimensional Rashba Hamiltonian \cite{rashba}
onto the basis of the two lowest uncoupled Rashba states, we
obtain the quasi one-dimensional Hamiltonian \cite{apl-cgd}
\begin{equation}
H=\left[
\begin{array}{cccc}
\varepsilon _{+}^{a}(k) & 0 & 0 & -\alpha d \\
0 & \varepsilon _{-}^{a}(k) & \alpha d & 0 \\
0 & \alpha d & \varepsilon _{+}^{b}(k) & 0 \\
-\alpha d & 0 & 0 & \varepsilon _{-}^{b}(k)%
\end{array}%
\right] ,  \label{ceq22}
\end{equation}%
where the interband coupling matrix element is $d\equiv \langle
\phi _{a}(y)|\partial /\partial y|\phi _{b}(y)\rangle $ and
$\phi_n(y)$ is the transverse channel wave function. Here we label
the uncoupled Rashba states by $n=a,b$ \cite{index}. The
Hamiltonian above gives rise to two sets of parabolic Rashba bands
for zero interband coupling $d=0$. These bands are sketched in
Fig. \ref{fig5c} (thin lines). Note that the uncoupled Rashba
bands cross. For positive $k$ vectors the crossing is at
$k_{c}=(\epsilon _{b}-\epsilon _{a})/2\alpha$. For non-zero
interband coupling $d\neq0$ the bands anti-cross near $k_c$ (see
thick lines); this follows from a straightforward diagonalization
of the 4x4 matrix in Eq. (\ref{ceq22}). We are interested here in
the \textit{weak} interband coupling limit. In addition, we
consider electron energies near the crossing; away from the
crossing the bands are essentially uncoupled and the problem
reduces to that of the previous section. In what follows, we adopt
a perturbative description for the energy bands near $k_c$ which
allows us to obtain analytical results.

\textit{``Nearly-free electron bands''.} In analogy to the usual
nearly-free electron approach in solids  \cite{am}, we restrict
the diagonalization of Eq. (\ref{ceq22}) to the 2x2 central block
which corresponds to the degenerate Rashba states crossing at
$k_c$
\begin{equation}
\tilde{H}=\left[
\begin{array}{cc}
\varepsilon _{-}^{a}(k) & \alpha d \\
\alpha d & \varepsilon _{+}^{b}(k)%
\end{array}
\right].  \label{ceq23}
\end{equation}
To lowest order we find
\begin{equation}
\varepsilon _{\pm }^{\mathrm{approx}}(k)=\frac{\hbar ^{2}k^{2}}{2m}+\frac{1}{2}%
\epsilon _{b}+\frac{1}{2}\epsilon _{a}\pm \alpha d.  \label{ceq24}
\end{equation}
The corresponding eigenvectors are the usual linear combination of
the \textit{zeroth order} degenerate states at the crossing
\begin{equation}
|\psi _{\pm }\rangle =\frac{1}{\sqrt{2}}\left[ |-\rangle _{a}\pm
|+\rangle _{b}\right], \label{ceq25}
\end{equation}%
where the ket sub-indices denote the respective (uncoupled) Rashba
channel [for simplicity, we omit the orbital part of the wave
functions in (\ref{ceq25})].

\subsubsection{Boundary conditions and spin injection near the crossing}
\label{two-band-injection}

Here we extend the analysis in Sec. \ref{one-band-injection} to
the case of two interband-coupled bands. We first determine the
$k$ points corresponding to the ``horizontal intersections'' near
the crossing at $k_c$, i.e., $k_{c1}$ and $k_{c2}$, see Fig.
\ref{fig5c}. We need these points since incoming spin-up electrons
will be primarily injected into those states (and also into $k_2$,
conservation of energy). By defining $k_{c1}=k_c - \Delta/2$ and
$k_{c2}=k_c+ \Delta/2$ and then imposing $\varepsilon
_{+}^{\mathrm{approx}}(k_{c1})=\varepsilon
_{-}^{\mathrm{approx}}(k_{c2})$ (assumed $\sim \varepsilon _{F}$)
we find,
\begin{equation}
\Delta =\frac{2m\alpha d}{\hbar
^{2}k_{c}}=2\frac{k_{R}}{k_{c}}d\mathrm{.} \label{ceq26}
\end{equation}

For a spin-up electron in the lowest wire state in the
``no-Rashba'' region (channel $a$), we can again write at $x=0$
\cite{unity}
\begin{eqnarray}
| &\uparrow & \rangle  e^{ikx}|_{x\rightarrow 0^{-}}=  \nonumber \\
&&\frac{1}{\sqrt{2}}\left\{ \frac{1}{\sqrt{2}}\left[ |\psi
_{+}\rangle e^{ik_{c1}x}+|\psi _{-}\rangle e^{ik_{c2}x}\right]
+|+\rangle _{a}e^{ik_{2}x}\right\} _{x\rightarrow 0^{+}},
\label{ceq26a}
\end{eqnarray}
in analogy to Eq. (\ref{ceq13}). Note that we only need to include
three intersection points in the above ``expansion'' since the
incoming spin-up electron is in channel $a$. Equation
(\ref{ceq26a}) satisfies the continuity of the wave function. The
boundary condition for the derivative of the wave function is also
satisfied provide that $\Delta/4\ll k_F$. This condition is
readily fulfilled for realistic parameters (Sec. \ref{estimates}).
Hence, fully spin-polarized injection into the Rashba region is
still possible in the presence of a weak interband coupling. Here
we are considering a fully spin-polarized injector so that the
intrinsic limitation due to the ``conductivity mismatch''
\cite{schmidt} is not a factor.

\textit{Generalized spin-rotated state at $x=L$.} Here again we
can easily determine the form of the state at the exit of the
Rashba region. For an incoming spin-up electron in the lowest band
of the wire, we find
\begin{equation}
\Psi _{\uparrow ,L}=\frac{1}{2}e^{i(k_{c}+k_{R})L}\left(
\begin{array}{c}
\cos (\theta _{d}/2)e^{-i\theta _{R}/2}+e^{i\theta _{R}/2} \\
-i\cos (\theta _{d}/2)e^{-i\theta _{R}/2}+ie^{i\theta _{R}/2} \\
-i\sin (\theta _{d}/2)e^{-i\theta _{R}/2} \\
\sin (\theta _{d}/2)e^{-i\theta _{R}/2}%
\end{array}%
\right). \label{ceq26d}
\end{equation}
A similar state holds for a spin-down incoming electron. The state
(\ref{ceq26d}) satisfies the boundary conditions at $x=L$ (again,
provided that $\Delta \ll 4k_F$. Equation (\ref{ceq26d})
essentially tells us that a weak s-o interband coupling gives rise
to an additional spin rotation (besides $\theta_R$) described by
the mixing angle $\theta_d=\theta_R d/k_c$. This extra modulation
enhances spin control in a Datta-Das spin-transistor geometry. In
Ref. \cite{apl-cgd} we show that the spin-resolved current in this
case is
\begin{equation}
I_{\uparrow ,\downarrow }=\frac{e}{h}eV[1\pm \cos (\theta
_{d}/2)\cos \theta _{R}], \label{new-dd}
\end{equation}
where $V$ is the source-drain bias.

\section{Novel Beam-splitter geometry with a local Rashba interaction}
\label{beam-splitter-rashba}

Figure \ref{bs-fig1} shows an schematic of our proposed
beam-splitter geometry with a local Rashba-active region of length
$L$ in lead 1. Below we discuss its scattering matrix in the
absence of interband coupling. In this case, each set of Rashba
bands can be treated independently.

\textit{Combined $\mathbf{s}$ matrices.} An electron entering the
system through port 1, first undergoes a unitary Rashba rotation
$\mathbf{U_{R}}$ in lead 1 then reaches the beam splitter which
either reflects the electron into lead 3 or transmits it into lead
4. This happens for electrons injected into either the first or
the second set of uncoupled Rashba bands. Since the Rashba spin
rotation is unitary, we can combine the relevant matrix elements
of the beam-splitter $\mathbf{s}$ matrix, connecting leads 1 and 3
($s_{14}=s_{41}$) and 1 and 4 ($s_{14}=s_{41}$), with the Rashba
rotation matrix $\mathbf{U_{R}}$ thus obtaining effective
\textit{spin-dependent} $2\times 2$  matrices of the form
$\mathbf{s^R_{13}}=\mathbf{s^R_{31}}=s_{13}\mathbf{U_{R}}$
A similar definition holds for
$\mathbf{s^R_{14}}=s_{41}\mathbf{U_{R}}=\mathbf{s^R_{41}}$. Note
also that $\mathbf{s_{23}}=\mathbf{s_{32}}=t\mathbf{1}$ and
$\mathbf{s_{24}}=\mathbf{s_{42}}=r\mathbf{1}$ since no Rashba
coupling is present in lead 2. All the other matrix elements are
zero. Hence the new effective beam-splitter $\mathbf{s}$ matrix
which incorporates the effect of the Rashba interaction in lead 1
reads
\begin{equation}
\mathbf{s}=\left(
\begin{array}{cccc}
\mathbf{0} & \mathbf{0} & \mathbf{s^R_{13}} & \mathbf{s^R_{14}} \\
\mathbf{0} & \mathbf{0} & \mathbf{s_{23}} & \mathbf{s_{24}} \\
\mathbf{s^R_{31}} & \mathbf{s_{32}} & \mathbf{0} & \mathbf{0} \\
\mathbf{s^R_{41}} & \mathbf{s_{42}} & \mathbf{0} & \mathbf{0}
\end{array}%
\right).  \label{ceq28}
\end{equation}
Note that incorporating the s-o effects directly into the
beam-splitter scattering matrix makes it spin dependent. The
Rashba interaction does not introduce any noise in lead 1. This is
so because the electron transmission coefficient through lead 1 is
essentially unity \cite{unity}; a quantum point contact is
noiseless for unity transmission.

\textit{Coupled Rashba bands.} The interband-coupled case can, in
principle, be treated similarly. However, we follow a different
simpler route to determine the shot noise in this case. We discuss
this in more detail in Sec. \ref{alternate-scheme}.

\section{Noise of entangled and spin-polarized electrons in
the presence of a local Rashba spin-orbit interaction}
\label{noise-rashba}

Starting from the noise definition in (Eq. \ref{ceq2}), we briefly
outline here the derivation of noise expressions for pairwise
electron states (entangled and unentangled) and spin-polarized
electrons (Secs. \ref{entanglers} and \ref{spinfilters}). For each
of these two cases, we present results with and without s-o
induced interband coupling.

\subsection{Shot noise for singlet and triplets}

\subsubsection{Uncoupled Rashba bands: single modulation $\theta_R$}
\label{noise-rashba-uncoupled}

To determine noise, we calculate the expectation value of the
noise operator (Eq. \ref{ceq2}) between pairwise electron states.
We have derived shot noise expressions for both singlet and
triplet states for a generic \textit{spin-dependent} $\mathbf{s}$
matrix. Our results quite generally show that unentangled triplets
and the entangled triplet display distinctive shot noise for
spin-dependent scattering matrices. Below we present shot noise
formulas for the specific case of interest here; namely, the
beam-splitter scattering matrix in the presence of a local Rashba
term [Eq. (\ref{ceq28})]. In this case, for singlet and triplets
defined along different quantization axes ($\hat{x}$ and $\hat{z}$
are equivalent directions perpendicular to the Rashba rotation
axis $-\hat{y}$), we find
\begin{equation}
S_{33}^{S}(\theta _{R})=\frac{2e^{2}}{h\nu }T(1-T)[1+\cos (\theta
_{R})\delta _{\varepsilon _{1},\varepsilon _{2}}],  \label{ceq31}
\end{equation}
\begin{equation}
S_{33}^{Te_y}(\theta _{R})=\frac{2e^{2}}{h\nu }T(1-T)[1-\cos
(\theta _{R})\delta _{\varepsilon _{1},\varepsilon _{2}}],
\label{ceq32}
\end{equation}
\begin{equation}
S_{33}^{Te_{z}}(\theta _{R})=S_{s-o}^{T_{u_y}}(\theta
_{R})=\frac{2e^{2}}{h\nu }T(1-T)(1-\delta _{\varepsilon
_{1},\varepsilon _{2}}),  \label{ceq33}
\end{equation}
and
\begin{equation}
S_{33}^{Tu_{\uparrow }}(\theta _{R})=S_{33}^{Tu_{\downarrow
}}(\theta _{R})=\frac{2e^{2}}{h\nu }T(1-T)[1-\cos ^{2}(\theta
_{R}/2)\delta _{\varepsilon _{1},\varepsilon _{2}}]. \label{ceq34}
\end{equation}
Equations (\ref{ceq32})--(\ref{ceq34}) clearly show that entangled
and unentangled triplets present distinct noise as a functions of
the Rashba phase. Note that for $\theta_R=0$, we regain the
formulas in Sec. \ref{earlier-results}. \vspace{0.5cm}
\begin{figure}[h]
\begin{center}
\epsfig{file=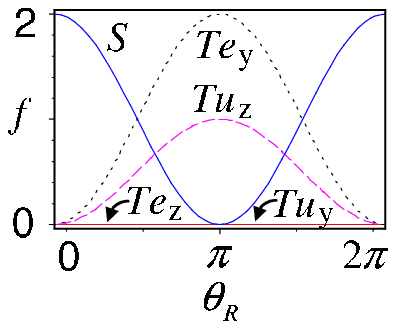,width=0.5\textwidth}
\end{center}
\vspace{-5mm} \caption{``Reduced'' Fano factor $f$ as a function
of the Rashba phase for singlet and triplets along different
quantization axes. Note that singlet and entangled triplet states
show \textit{continuous} bunching and antibunching behaviors as
$\theta_R$ is increased. Unentangled triplets display distinctive
noise for a given polarization and for different polarizations.
Adapted from Ref. \cite{prl-cgd}.} \label{fano-b}
\end{figure}

Figure \ref{fano-b} shows the ``reduced'' Fano factor
$f=F/2T(1-T)$, $F=S_{33}/eI$ (here $I=e/h\nu$), as a function of
the Rashba angle $\theta_R$ for the noise expressions
(\ref{ceq31})--(\ref{ceq34}). We clearly see that singlet and
triplet pairs exhibit distinct shot noise in the presence of the
s-o interaction. The singlet $S$ and entangled (along the Rashba
rotation axis $\hat y$) triplet $Te_y$ pairs acquire an
oscillating phase in lead 1 thus originating intermediate degrees
of bunching/antibunching (solid and dotted lines, respectively).
Triplet states (entangled and unentangled) display distinctive
noise as a function of the Rashba phase, e.g., $T{e_y}$ is noisy
and $T{u_y}$ is noiseless. Hence entangled and unentangled
triplets can also be distinguished via noise measurements. Note
that for $\theta_R=0$ all three triplets exhibit identically zero
noise [see Eq.(\ref{ceq7})].

\subsubsection{Interband-coupled Rashba bands: additional modulation $\theta_d$}
\label{alternate-scheme}

Here we determine noise for injected pairs with energies near the
crossing $\varepsilon(k_c)$ using an alternate scheme. We
calculate the relevant expectation values of the noise by using
pairwise states defined from the generalized spin-rotated state in
Eq. (\ref{ceq26d}) and its spin-down counterpart. Since these
states already incorporate all the relevant effects (Rashba
rotation and interband mixing), we can calculate noise by using
the ``bare'' beam splitter matrix elements, generalized to account
for two channels. The beam-splitter does not mix transverse
channels; hence this extension is trivial, i.e., block diagonal in
the channel indices. This approach was first developed in Ref.
\cite{jsc-cgd}.

\textit{Rashba-evolved pairwise electron states.} The portion of
an electron-pair wave function ``propagating'' in lead 1 undergoes
the effects of the Rashba interaction: ordinary precession
$\theta_R$ and additional rotation $\theta_d$. Using Eq.
(\ref{ceq26d}) (and its spin-down counterpart) we find the
following states
\begin{eqnarray}
|S/Te_z\rangle_L &=&\frac{1}{2}[\cos (\theta _{d}/2)e^{-i\theta _{R}/2}+e^{i\theta _{R}/2}]%
\frac{|\uparrow \downarrow \rangle _{aa}\mp |\downarrow \uparrow
\rangle
_{aa}}{\sqrt{2}}+ \nonumber \\
&&\frac{1}{2}[-i\cos (\theta _{d}/2)e^{-i\theta _{R}/2}+ie^{i\theta _{R}/2}]\frac{%
|\downarrow \downarrow \rangle _{aa}\pm |\uparrow \uparrow \rangle
_{aa}}{\sqrt{2}}+ \nonumber \\
&&\frac{1}{2}[-i\sin (\theta _{d}/2)e^{-i\theta
_{R}/2}]\frac{|\uparrow \downarrow
\rangle _{ba}\pm |\downarrow \uparrow \rangle _{ba}}{\sqrt{2}}+ \nonumber \\
&&\frac{1}{2}[\sin (\theta _{d}/2)e^{-i\theta
_{R}/2}]\frac{|\downarrow \downarrow \rangle _{ba}\mp |\uparrow
\uparrow \rangle _{ba}}{\sqrt{2}} \label{ceq36}.
\end{eqnarray}
The notation $|T{e_z}\rangle_L$ and $|S\rangle_L$ emphasizes the
type of injected pairs (singlets or triplets at $x=0$) propagating
through the length $L$ of the Rashba-active region in lead 1.
Similar expressions hold for $|Tu_{\uparrow,\downarrow} \rangle
_{L}$. In addition, we use the shorthand notation $|\downarrow
\uparrow \rangle _{ba} \equiv |\downarrow_{1b} \uparrow_{2a}
\rangle$, denoting a pair with one electron in channel \textit{b}
of lead 1 and another in channel \textit{a} of lead 2. Here we
consider incoming pairs with $\hat z$ polarizations only. Despite
the seemingly complex structure of the above pairwise states, they
follow quite straightforwardly from the general state $\Psi
_{\uparrow ,L}$ in (\ref{ceq26d}) (and its counterpart $\Psi
_{\downarrow ,L}$). For instance, the unentangled triplet
$|Tu_\uparrow \rangle _{L}$ is obtained from the tensor product
between $\Psi _{\uparrow ,L}$ [which describes as electron
crossing lead 1 (initially spin up and in channel \textit{a})] and
a spin-up state in channel $a$ of lead 2: $|Tu_\uparrow \rangle
_{L} = |\Psi _{\uparrow ,L} \rangle \bigotimes |\uparrow
\rangle_{2a}$.

\textit{Noise.} We can now use the above states to determine shot
noise at the zero frequency, zero temperature, and zero applied
bias. Using the shot-noise results of Sec. \ref{earlier-results}
(trivially generalized for two channels), we find for the noise in
lead 3
\begin{eqnarray}
S_{33}^{Tu_{\uparrow }}(\theta _{R},\theta _{d})
&=&S_{33}^{Tu_{\downarrow }}(\theta _{R},\theta
_{d})=\frac{2e^{2}}{h\nu }
T(1-T)\times   \nonumber \\
&& \left[ 1-\frac{1}{2}\left( 1 +\cos \frac{\theta _{d}}{2} \cos
\theta _{R}-\frac{1}{2}\sin ^{2}\frac{\theta _{d}}{2}\right)
\delta _{\varepsilon _{1},\varepsilon _{2}}\right], \qquad
\label{ceq37}
\end{eqnarray}
\begin{equation}
S_{33}^{Te_{z}}(\theta _{R},\theta _{d})=\frac{2e^{2}}{h\nu }T(1-T)\left[ 1-\frac{%
1}{2}\left( \cos ^{2}\frac{\theta _{d}}{2} +1\right) \delta
_{\varepsilon _{1},\varepsilon _{2}}\right] ,  \label{ceq38}
\end{equation}
and
\begin{equation}
S_{33}^{S}(\theta _{R},\theta _{d})=\frac{2e^{2}}{h\nu
}T(1-T)\left[ 1+\left( \cos \frac{\theta _{d}}{2}\cos \theta
_{R}\right) \delta _{\varepsilon _{1},\varepsilon _{2}}\right].
\label{ceq39}
\end{equation}
Equations (\ref{ceq37})-(\ref{ceq39}) describe shot noise only for
injected pairs with energies near the crossing, say, within
$\alpha d$ of $\varepsilon(k_c)$. Away from the crossing or for
$d=0$, the above expressions reduce to those of Sec.
\ref{noise-rashba-uncoupled}.
We can also define ``reduced'' Fano factors as before;
the interband mixing angle $\theta_d$ further modulates the Fano
factors. For conciseness, we present the angular dependence of the
Fano factors in the next section.

\subsection{Shot noise for spin-polarized electrons}
\label{noise-spin-pol}

We have derived a general shot noise formula for the case of
spin-polarized sources by performing the ensemble average in Eq.
(\ref{ceq2}) over appropriate thermal reservoirs. The resulting
expression corresponds to the standard Landauer-B\"uttiker formula
for noise with spin-dependent $\mathbf{s}$ matrices. Below we
present results for the specific beam-splitter $\mathbf{s}$ matrix
in (\ref{ceq28}).

\subsubsection{Uncoupled-band case: single modulation $\theta_R$}

\begin{figure}[ht]
\begin{center}
\epsfig{file=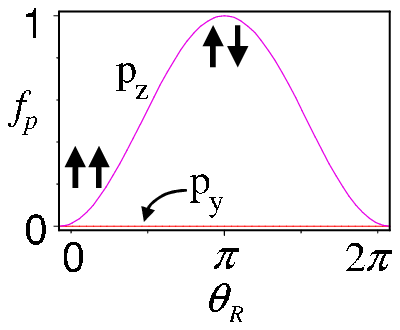,width=0.5\textwidth}
\end{center}
\vspace{-5mm} \caption{``Reduced'' Fano factor for fully
spin-polarized ($p=1$) incoming beams in leads 1 and 2 as a
function of the Rashba phase. Polarizations along two distinct
quantization axes are shown ($p_z$ and $p_z$). For spin injection
along the Rashba rotation axis ($-\hat y$), no precession occurs
in lead 1 and shot noise is identically zero (Pauli principle).
Spin-polarized carriers injected along $\hat z$ undergo precession
and hence exhibit shot noise. Adapted from Ref. \cite{prl-cgd}.}
\label{fano-a}
\end{figure}

For incoming leads with a degree of spin polarization $p$ and for
the scattering matrix (\ref{ceq28}), we find at zero temperatures
\begin{equation}
S_{33}^{p} (\theta_R) =  2e I  T(1-T)p \sin^{2}\frac{\theta
_{R}}{2}, \label{ceq43}
\end{equation}
where $I=2e^2V/[h(1+p)]$ is the average current in lead 3. The
``reduced'' Fano factor corresponding to Eq. (\ref{ceq43}) is
$f_{p}  =  p \sin^{2}(\theta _{R}/2)$. Figure \ref{fano-a} shows
$f_{p}$ as a function of the Rashba angle $\theta_R$. For spin
polarized injection along the Rashba rotation axis ($-\hat y$) no
noise results in lead 3. This is a consequence of the Pauli
exclusion principle in the leads. Spin-polarized currents with
polarization perpendicular to the Rashba axis exhibit sizable
oscillations as a function of $\theta_R$. Full shot noise is
obtained for $\theta_R=\pi$ since the spin polarization of the
incoming flow is completely reversed within lead 1. \vspace{1.0cm}

\textit{Probing/detecting spin-polarized currents.} Since
\textit{unpolarized} incoming beams in lead 1 and 2 yield zero
shot noise in lead 3, the results shown in Fig. \ref{fano-a}
provide us with an interesting way to \textit{detect}
spin-polarized currents via their noise. In addition, noise
measurements should also allow one to probe the direction of the
spin-polarization of the injected current.

\textit{Measuring the s-o coupling.} We can express the s-o
coupling constant in terms of the reduced Fano factor. For a fully
spin-polarized beam ($p=1$), we have
\begin{equation}
\alpha =\frac{\hbar ^{2}}{m^{\ast }L}\arcsin \sqrt{f_{p}}.
\label{ceq46}
\end{equation}
Equation (\ref{ceq46}) provides a direct means of extracting the
Rashba s-o coupling $\alpha$ via shot noise measurements. We can
also obtain a similar expression for $\alpha$ from the unentangled
triplet noise formula (\ref{ceq34}).

\subsubsection{Interband-coupled case: extra modulation $\theta_d$}

The calculation in the previous section can be extended to the
interband-coupled case for electrons impinging near the anti
crossing of the bands [$\sim \varepsilon(k_c)$]. Here we present a
simple ``back-of-the-envelope'' derivation of the the shot noise
for the fully spin-polarized current case $(p=1)$ from that of the
spin-up unentangled triplet Eq. (\ref{ceq37}). Here we imagine
that the spectrum of the triplet $Tu_{\uparrow }$ forms now a
continuum and integrate its noise expression (after making
$\varepsilon_1=\varepsilon_2$) over some energy range to obtain
the noise of a spin-polarized current. Assuming $T$ constant in
the range ($\varepsilon_F,\varepsilon_F+eV$), we find to linear
order in $eV$
\begin{equation} S_{33}^{\uparrow }(\theta _{R},\theta _{d})
= eI T(1-T) \left( 1 -\cos \frac{\theta _{d}}{2} \cos \theta
_{R}+\frac{1}{2}\sin ^{2}\frac{\theta _{d}}{2}\right).
\label{ceq49}
\end{equation}

\begin{figure}[th]
\begin{center}
\epsfig{file=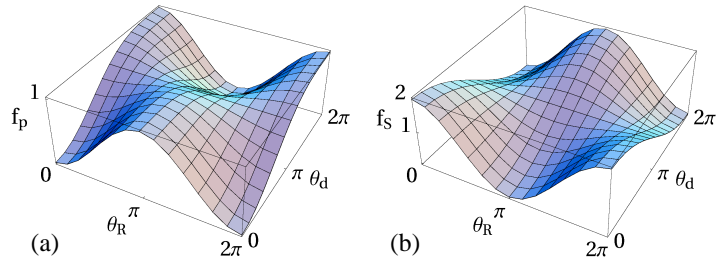,width=0.80\textwidth}
\end{center}
\par
\vspace{-8mm} \caption{Reduced Fano factors $f=f_{p}$ (a) and
$f_S$ (b), for fully spin-polarized ($p=1$, $\hat z$ direction)
incoming electrons and for singlet pairs, respectively, as a
function of the Rashba angle $\theta_R$ and the interband mixing
angle $\theta_d$. The additional phase $\protect\theta _{d}$ can
significantly alter the noise characteristics.} \label{fano3d}
\end{figure}

Figures \ref{fano3d}(a) and \ref{fano3d}(b) illustrate the angular
dependencies of the reduced Fano factors for both the
spin-polarized case Eq. (\ref{ceq49}) and that of the singlet Eq.
(\ref{ceq39}). Note that the further modulation $\theta_d$ due to
interband mixing can drastically change the noise for both
spin-polarized and entangled electrons. For the singlet pairs, for
instance, it can completely reverse the bunching/antibunching
features. Hence further control is gained via $\theta_d$ which
can, in principle, be tuned independently of $\theta_R$ (see Sec.
\ref{estimates}).

\subsection{Realistic parameters: estimates for $\theta_R$ and $\theta_d$.}
\label{estimates}

We conclude this section by presenting some estimates for the
relevant spin-rotation angles $\theta_R$ and $\theta_d$ for
realistic system parameters. Let us assume, for the sake of
concreteness, an infinite confining potential of width $w$. In
this case, the transverse wire modes in absence of the Rashba
interaction are quantized with energies $\epsilon_n=\hbar^2\pi^2
n^2/(2m^\ast w^2)$. Let us now set
$\epsilon_b-\epsilon_a=3\hbar^2\pi^2/(2m^\ast w^2)=16\epsilon_R$
which is a ``reasonable guess''. Since
$\epsilon_R=m^\ast\alpha^2/2\hbar^2$, we find
$\alpha=(\sqrt{3}\pi/4)\hbar^2/m^\ast w^2=3.45\times 10^{-11}$ eVm
\cite{gatecontrol} (which yields $\epsilon_R\sim0.39$ meV) for
$m^\ast=0.05m_0$ and $w=60$ nm. For the above choice of
parameters, the energy at the crossing is
$\epsilon^a_-(k_c)=\epsilon^b_+(k_c)=\epsilon(k_c)=24\epsilon_R\sim
9.36$ meV. Electrons with energies around this value are affected
by the s-o interband coupling, i.e., they undergo the additional
spin rotation $\theta_d$. The relevant wave vector at the crossing
is $k_c=8\epsilon_R/\alpha$. Assuming the $L=69$ nm for the length
of the Rashba channel, we find $\theta_R=\pi$ and
$\theta_d=\theta_R d/k_c\sim \pi/2$ since $d/k_c=2/(3k_Rw)$ and
$k_Rw=\sqrt{3}\pi/4\sim4/3$ for
$\epsilon_b-\epsilon_a=16\epsilon_R$ which implies $d/k_c\sim
0.5$. The preceding estimates are conservative. We should point
out that both $\theta_R$ and $\theta_d$ can, in principle, be
varied independently via side gates. It should also be possible to
``over rotate'' $\theta_R$ (say, by using a larger $L$) and hence
increase $\theta_d$. As a final point we note that $\Delta/4k_F
\sim 0.05\ll 1$ [$k_F$ is obtained by making
$\varepsilon_F=\hbar^2k^2_F/2m^\ast=\epsilon(k_c)$] which assures
the validity of the boundary condition for the velocity operator.

\section{Relevant issues and outlook}
\label{rel-issues}

\textit{Relevant time scales.} Typical parameters for a
finite-size electron beam splitter (tunnel coupled to reservoirs)
defined on a GaAs 2DEG are: a device size $L_0\sim 1$ $\mu$m, a
Fermi velocity in the range $v_F\sim 10^4 - 10^5$ m/s and an
orbital coherence length of $\sim 1$ $\mu$m \cite{bs}. These
values lead to traversal times $\tau_t = L_0/v_F$ in the range
$\sim 10-100$ ps; these are lower bounds for the actual dwell time
$\tau_{\rm dwell}\sim 1/\gamma_R$ of the electrons in the beam
splitter, where $\gamma_R$ is the tunnelling rate from the leads
of the beam splitter to the reservoirs. Hence the electrons keep
their orbital coherence across the beam-splitter at low
temperatures. Moreover, long spin dephasing times in
semiconductors ($\sim 100$ ns for bulk GaAs \cite{Kik97}) should
allow the propagation of entangled electrons without loss of spin
coherence.

For the noise calculation with entangled/unentangled pairs, we
have assumed discrete energy levels in the incoming leads. A
``particle-in-a-box'' estimate of the level spacing
$\delta\varepsilon$ due to longitudinal quantization of the beam
splitter leads yields $\delta\varepsilon \sim \hbar v_F/L_0\sim
0.01-0.1$ meV. The relevant broadening of these levels is given by
the coupling $\gamma_R \ll \delta \varepsilon $, which justifies
the discrete level assumption. Here we take $\gamma_R \lesssim
\gamma \sim 1$ $\mu$eV,  where $\gamma$ is the tunnelling rate
from the entangler to the beam splitter (Sec. \ref{entanglers}).
In addition, the stationary state description we use requires that
the electrons have enough time to ``fill in'' the extended states
in the beam splitter before they leave to the reservoirs:
$\tau_{\rm dwell} \gtrsim \tau_{\rm inj} \sim 1/\gamma$. Here
$\tau_{\rm inj}$ is the injection time from the entangler to the
beam splitter. To have well separated pairs of entangled
electrons, we also need $\tau_{\rm delay} < \tau_{\rm pairs} \sim$
ns (Sec. \ref{entanglers}), where $\tau_{\rm delay}$ ($\sim $ ps)
is the time delay  between two entangled electrons of the same
pair, and $\tau_{\rm pairs}$ ($\sim $ ns) is the time separation
between two subsequent pairs.

\textit{Interactions in the beam splitter.} For entangled
electrons it would be advantageous to reduce electron-electron
interaction in the beam splitter, which is the main source of
orbital decoherence at low temperatures. This could be achieved by
depleting the electron sea in the beam splitter, e.g., by using
the lowest channel in a quantum point contact. A further
possibility is to use a superconductor for the beam splitter
\cite{leo}. A superconductor would have the advantage that the
entangled electrons could be injected into the empty quasiparticle
states right at their chemical potential. Because of the large gap
$\Delta$ between these states and the condensate, the injected
electron cannot exchange energy (nor spin) with the underlying
condensate of the superconductor.

An alternative way to detect entangled pairs would be to use a
superconductor as an analyzer: arriving entangled (spin-singlet)
pairs can enter the superconductor whereas any triplet state is
not allowed. Thus, the current of entangled pairs is larger than
otherwise.

\section{Noise of a double QD near the Kondo regime}

\newcommand{\T}[0]{{\rm T}}

Spin-flip processes in a spin $1/2$ quantum dot attached to leads
result in a
 renormalization of the single-particle transmission coefficient ${\rm T}$,
 giving rise to the Kondo effect~\cite{GRNgL}
 below the Kondo temperature $T_K$.
Theoretical studies on shot noise in this system are
 available~\cite{MeirGolub}--\cite{SchillerHershfield},
 and show that the noise $S$ obeys qualitatively
 the same formula as for noninteracting electrons but with a
 renormalized ${\rm T}$.
Here, we consider a system where the spin fluctuations (that are
enhanced
 near the Kondo regime) strongly affect the noise,
 resulting in some cases in super-Poissonian noise -- a result which cannot
 be obtained from the ``non-interacting'' formula.

\begin{figure}[h]
\centerline{\psfig{file=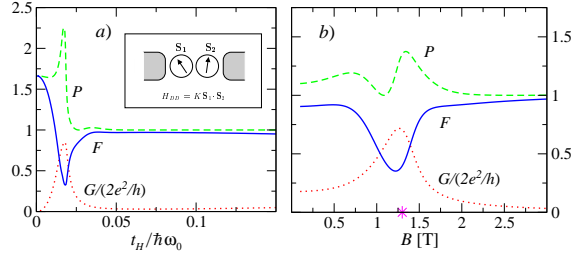,width=7.5cm}} \caption{
{\em a}) Linear conductance $G$ (dotted line),
 Fano factor (solid line), and the factor $P$
 (dashed line), in vicinity of the 2IKM critical point.
Inset: DD setup. {\em b}) Similar to ({\em a}), but in the
vicinity of
 the singlet-triplet Kondo effect (``*'' denotes $K=0$).}
\label{DDKondo2}
\end{figure}

We consider two lateral quantum dots (DD),
 connected in series between two metallic leads via tunnel contacts, see
 inset of Fig.~\ref{DDKondo2}{\em a}.
The dots are tuned into the Coulomb blockade regime,
 each dot having a spin $1/2$ ground state.
The low energy sector of the DD consists of a singlet $|S\rangle$
 and a triplet
 $|T\rangle\equiv\left\{|T_+\rangle,|T_0\rangle,|T_-\rangle\right\}$,
 with the singlet-triplet splitting $K$.
The Kondo effect in this system has been studied
 extensively~\cite{GolovachLoss}--\cite{AnoEto}.
Two peculiar features in the linear conductance $G$ have been
found:
 a peak in $G$ {\em vs} the inter-dot tunnel coupling $t_H$
 (see Fig.~\ref{DDKondo2}{\em a}),
 revealing the non-Fermi-liquid critical point of
 the two-impurity Kondo model (2IKM)~\cite{ALJ}; and
 a peak in $G$ {\em vs} an applied perpendicular magnetic field $B$
 (see Fig.~\ref{DDKondo2}{\em b}),
 as a result of the singlet-triplet Kondo effect at
 $K=0$~\cite{GolovachLoss}.

The problem of shot noise in DDs with Kondo effect is rather
involved. Here we propose a phenomenological approach. For bias
$\Delta\mu\gg T_K,K$,
 the scattering problem can be formulated in terms
 of the following scattering matrix
\begin{eqnarray}
{\rm s}&=& \left(
\begin{array}{cc}
r_S & t_S\\
t_S & r_S
\end{array}
\right)\left|S\right\rangle\left\langle S\right|+ \left(
\begin{array}{cc}
r_T & t_T\\
t_T & r_T
\end{array}
\right)\left|T\right\rangle\left\langle T\right|\nonumber\\
&&+\left(
\begin{array}{cc}
r_{TS} & t_{TS}\\
t_{TS} & r_{TS}
\end{array}
\right)\left|T\right\rangle\left\langle S\right|+ \left(
\begin{array}{cc}
r_{ST} & t_{ST}\\
t_{ST} & r_{ST}
\end{array}
\right)\left|S\right\rangle\left\langle T\right|, \label{Smatrix}
\end{eqnarray}
 where $t_{i(j)}$ and $r_{i(j)}$ are the transmission and reflection
 amplitudes.
The spin fluctuations in the DD cause fluctuations in the
transmission
 through the DD.
The dominant mechanism is qualitatively
 described by the following stochastic model
\begin{equation}
f(t)=\left[f_1(t)\left(1-F(t)\right)+f_2(t)F(t)\right]
\left(1-\left|\dot{F}(t)\right|\right)+f_3(t)\left|\dot{F}(t)\right|,
\end{equation}
where $f_i(t)=0,1$ is a white noise ($i=1,2,3$) with $\langle
f_i(t)\rangle=\bar{f}_i$ and $\langle
f_i(t)f_i(0)\rangle-\bar{f}_i^2=\bar{f}_i(1-\bar{f}_i)
\delta(t/\Delta t)$, and $F(t)=0,1$ is a telegraph noise with
$\bar{F}=\beta/(1+\beta)$ and $\langle
F(t)F(0)\rangle-\bar{F}^2=\beta\exp(-ct)/(1+\beta)^2$, for $t\geq
0$. In this model, the time $t$ is discretized in intervals of
 $\Delta t=h/2\Delta\mu$.
The derivative $\dot{F}(t)$ takes values $0,\pm 1$. The function
$f_{1(2)}(t)$ describes tunnelling through
 the DD, with the DD staying in the singlet (triplet) state, while
 $f_3(t)$ describes tunnelling accompanied
 by the DD transition between singlet and triplet.
The relation to formula (\ref{Smatrix})
 is given by:
$\bar{f}_1=|t_S|^2=1-|r_S|^2$, $\bar{f}_2=|t_T|^2=1-|r_T|^2$, and
$f_3=|t_{ST}|^2/\left(|t_{ST}|^2+|r_{ST}|^2\right)=
|t_{TS}|^2/\left(|t_{TS}|^2+|r_{TS}|^2\right)$. The telegraph
noise is described by two parameters: $\beta=w_{12}/w_{21}$ and
$c=w_{12}+w_{21}$, where $w_{ij}$ is the probability to go from
$i$ to $j$.

The quantity of interest is the Fano factor $F=S/e|I|$. For a
single-channel non-interacting system, one has $F=1-{\rm T}$. In
order to show the effect of
 interaction, we introduce the factor $P=F/(1-\T)$.
The noise power at
 zero frequency is then given by $S=2eI_{\rm imp}\T(1-\T)P$, where
 $I_{\rm imp}=2e\Delta\mu/h$.
For the average transmission probability we obtain
\begin{equation}
\T\equiv\langle f\rangle=\frac{\bar{f}_1+\beta\bar{f}_2}{1+\beta}+
\frac{\beta c\Delta t}{(1+\beta)^2}\left(2\bar{f}_3-
\bar{f}_1-\bar{f}_2\right).
\end{equation}
The noise can be calculated as $S=2eI_{\rm imp}S_f$, with
$S_f=\T(1-\T)+\Delta S_f$, where
\begin{eqnarray}
\Delta S_f&=& \frac{2\beta}{(1-q)(1+\beta)^2}
\left\{q(\bar{f}_1-\bar{f}_2)^2+ \frac{c\Delta
t(\bar{f}_1-\bar{f}_2)}{(1+\beta)}\times
\right.\nonumber\\
&& \left[\bar{f}_3(\beta-1)(q+1)+\bar{f}_1(1-\beta
q)+\bar{f}_2(q-\beta)\right]+
\nonumber\\
&& \left.\frac{(c\Delta t)^2}{4}
\left[\left(2\bar{f}_3-\bar{f}_1-\bar{f}_2\right)^2-
\left(\bar{f}_1-\bar{f}_2\right)^2\right]\right\},
\end{eqnarray}
with $q=\exp(-c\Delta t)$. The factor $P$ is then given by
$P=1+\Delta S_f/(\T-\T^2)$. Deviations of $P$ from $P=1$ show the
effect of interactions in the DD. We plot the Fano factor and the
factor $P$ for a DD on Fig.~\ref{DDKondo2}. The results show that
the spin fluctuations affect the shot noise in the regions where
$K\lesssim T_K$. A peculiar feature in $P$ is found both at the
2IKM critical point (Fig.~\ref{DDKondo2}{\em a}) and at the point
of the singlet-triplet Kondo effect (Fig.~\ref{DDKondo2}{\em b}).

For $\Delta\mu\ll T_K$ the DD spin is screened, and correlations
between
 two electrons passing through the DD
 occur only via virtual excitations
 of the Kondo state.
The shot noise is expected to qualitatively
 obey the non-interacting formula with
 the renormalized $\T$.

\section{Summary}

We presented our recent works on shot noise for spin-entangled
electrons and spin-polarized currents in novel beam splitter
geometries. After a detailed description of various schemes
(``entanglers'') to produce entangled spin states, we calculated
shot noise within the scattering approach for a beam splitter with
and without a local s-o interaction in the incoming leads. We find
that the s-o interaction significantly alters the noise.
Entangled/unentangled pairs and spin-polarized currents show
sizable shot noise oscillations as a function of the Rashba phase.
Interestingly, we find an additional phase modulation due to s-o
induced interband coupling in leads with two channels. Shot noise
measurements should allow the identification/characterization of
both entangled and unentangled pairs as well as spin-polarized
currents. Finally, we find that the s-o coupling constant $\alpha$
is directly related to the Fano factor; this offers an alternative
means of extracting $\alpha$ via noise.

This work was supported by NCCR\ Nanoscience, the Swiss NSF,
DARPA, and ARO.


\end{article}

\begin{thebibliography}{}
\bibitem{cost}
This reduction factor of the current $I_{2}$ compared to the
resonant current $I_{1}$ reflects the energy cost in the virtual
states when two electrons tunnel via the same QD into the same
Fermi lead and are given by $U$ and/or $\Delta$. Since the
lifetime broadenings $\gamma_{1}$ and $\gamma_{2}$ of the two QDs
1 and 2 are small compared to $U$ and $\Delta$ such processes are
suppressed.
\bibitem{LLfootnote}
The interaction dependent constants $A_{b}$ are of order one for
not too strong interaction between electrons in the LL but are
decreasing when interaction in the LL-leads  is increased
\cite{P8RL}. Therefore in the case of substantially strong
interaction as it is present in metallic carbon nanotubes, the
pre-factors $A_{b}$ can help in addition to suppress $I_{2}$.
\bibitem{P21electronbunching}
Since $\gamma_{\rho -}>\gamma_{\rho +}$, it is more probable that
two electrons coming from the same Cooper pair travel in the same
direction than into different directions when injected into the
same LL-lead.
\bibitem{spinexc}
In order to have exclusively singlet states as an input for the
beamsplitter setup, it is important that the LL-leads return to
their spin ground-state after the injected electrons have
tunnelled out again into the Fermi leads. For an infinite LL, spin
excitations are gapless and therefore an arbitrary small bias
voltage $\mu$ between the SC and the Fermi liquids gives rise to
spin excitations in the LL. However, for a realistic finite size
LL (e.g. a nanotube), spin excitations are gapped on an energy
scale $\sim\hbar v_{F}/L$, where $L$ is the length of the LL.
Therefore, if $k_{B}T,\mu<\hbar v_{F}/L$ only singlets can leave
the LL again to the Fermi leads, since the total spin of the
system has to be conserved. For metallic carbon nanotubes, the
Fermi velocity is $\sim 10^{6}{\rm m/s}$, which gives an
excitation gap of the order of a few meV for $L\sim$ $\mu$m; this
is large enough for our regime of interest.
\bibitem{charging}This symmetric setup of the charging energy \(U\) is obtained when the gate voltages
are tuned such that the total Coulomb charging energies in \dc are
equal with zero or two electrons.
\bibitem{patrik-sf-memory} In addition, for fully spin-polarized leads the
device can act as a single spin memory with read-in and read-out
capabilities if the dot is subjected to a ESR source.
\bibitem{patrik-sf-filter} This is true as long as the Zeeman splitting in the leads is much
smaller than their Fermi energies.
\bibitem{factor-of-two} Our noise definition here differs by a
factor of two from that in the review article by Blanter and
B\"uttiker (Ref. \cite{blanter-buttiker}); these authors define
their power spectral density of the noise with a coefficient two
in front (see definition following Eq. (49)  and footnote 4 in
Ref. \cite{blanter-buttiker}). We use a standard Fourier transform
(no factor of two in front) to define the noise spectral density.
\bibitem{discrete} For a discrete energy spectrum we need to insert a density-of-states factor $\nu$
in the current and noise definitions; see Ref. \cite{gde}.
\bibitem{fu} Note that the uncorrelated-beam
case here refers to a beam splitter configuration with only one of
the incoming leads ``open''. This is an important point since a
beam splitter is noiseless for (unpolarized) uncorrelated beams in
both incoming leads.
\bibitem{bc} Note that the velocity operator is not diagonal in the
presence of the Rashba interaction.
\bibitem{index} In the absence of the s-o interaction, we assume the wire
has two sets of spin-degenerate parabolic bands for each $k$
vector. In the presence of s-o interaction but neglecting s-o
induced interband coupling, there is a one-to-one correspondence
between the parabolic bands with no spin orbit and the Rashba
bands; hence they can both be labelled by the same indices.
\end{thebibliography}
\end{document}